\documentclass[10pt,twocolumn]{IEEEtran}
\usepackage{times}
\usepackage{latexsym}
\usepackage[usenames, dvipsnames]{color}
\usepackage{stfloats}
\usepackage{footnote}
\usepackage{graphicx,amsfonts,amsmath,amssymb,array,url}
\usepackage{amsthm}
\usepackage{extarrows}
\usepackage{ifpdf}
\usepackage{cite}
\usepackage{algorithmic}
\usepackage{epsfig}
\usepackage{epstopdf}
\usepackage{mathtools}
\usepackage{caption}
\usepackage{subcaption}
\usepackage[utf8]{inputenc}
\usepackage{epstopdf}
\usepackage{setspace}
\usepackage{multicol}
\definecolor{green}{rgb}{0.796,0.948,0.816}
\theoremstyle{remark} 
\newtheorem{Term0_MRC}{Proposition}
\newtheorem{Term1_MRC}[Term0_MRC]{Proposition}
\newtheorem{Term2_MRC}[Term0_MRC]{Proposition}
\newtheorem{Term3_MRC}[Term0_MRC]{Proposition}
\newtheorem{Term4_MRC}[Term0_MRC]{Proposition}

\newtheorem{Term1_ZF}[Term0_MRC]{Proposition}
\newtheorem{Trace Connection}[Term0_MRC]{Proposition}
\newtheorem{MMSE_error}[Term0_MRC]{Proposition}
\newtheorem{Lemma1_MRC}{Lemma}
\newtheorem{Lemma2_MRC}[Lemma1_MRC]{Lemma}
\newtheorem{Lemma3_MRC}[Lemma1_MRC]{Lemma}
\newtheorem{Lemma4_MRC}[Lemma1_MRC]{Lemma}
\newtheorem{Lemma5_MRC}[Lemma1_MRC]{Lemma}
\newtheorem{Corollary1_MRC}{Corollary}

\newcommand{\tc}{{\tau}_c}
\newcommand{\ts}{{\tau}_s}

\newcommand{\Rch}{\hat{\mathbf{R}}_c}
\newcommand{\Fc}{\mathbf{F}_c}
\newcommand{\Rc}{\mathbf{R}_c}
\newcommand{\Fd}{\mathbf{F}_d}
\newcommand{\yc}{\mathbf{y}_c}
\newcommand{\nc}{\mathbf{n}_c}

\newcommand{\go}{\mathbf{g}_1}

\newcommand{\Ypt}{\mathbf{Y}_{pt}}
\newcommand{\Ypr}{\mathbf{Y}_{pr}}
\newcommand{\Yo}{\mathbf{Y}_1}

\newcommand{\Npr}{\mathbf{N}_{pr}}
\newcommand{\Npt}{\mathbf{N}_{pt}}
\newcommand{\No}{\mathbf{N}_1}

\newcommand{\Phio}{\mathbf{\Phi}_1}
\newcommand{\Phit}{\mathbf{\Phi}_2}
\newcommand{\tp}{\tau_p}

\newcommand{\U}{\mathbf{U}}

\newcommand{\Q}{\mathbf{Q}}
\newcommand{\Ro}{\mathbf{R}_1}
\newcommand{\Uro}{\mathbf{U}_{R_1}}
\newcommand{\Sro}{\mathbf{\Lambda}_{R_1}}
\newcommand{\Rt}{\mathbf{R}_2}

\newcommand{\Fo}{\mathbf{F}_1}
\newcommand{\Uo}{\mathbf{U}_{1}}
\newcommand{\Ueo}{\mathbf{U}_{e_1}}
\newcommand{\Ut}{\mathbf{U}_{2}}
\newcommand{\Uet}{\mathbf{U}_{e_2}}
\newcommand{\Ufo}{\mathbf{U}_{F_1}}
\newcommand{\Sfo}{\mathbf{\Sigma}_{F_1}}
\newcommand{\Vfo}{\mathbf{V}_{F_1}}
\newcommand{\Ft}{\mathbf{F}_2}

\newcommand{\Ho}{\mathbf{H}_1}
\newcommand{\Ht}{\mathbf{H}_2}
\newcommand{\Go}{\mathbf{G}_1}
\newcommand{\Gt}{\mathbf{G}_2}

\newcommand{\W}{\mathbf{W}}
\newcommand{\G}{\mathbf{G}}
\newcommand{\Eo}{\mathbf{E}_1}
\newcommand{\Et}{\mathbf{E}_2}

\newcommand{\nr}{\mathbf{n}_R}

\newcommand{\Hoh}{\tilde{\mathbf{H}}_1}

\newcommand{\Goh}{\hat{\mathbf{G}}_1}
\newcommand{\Gth}{\hat{\mathbf{G}}_2}

\newcommand{\Heo}{\tilde{\mathbf{H}}_{e1}}

\newcommand{\gm}{\mathbf{g}_{m}}

\newcommand{\gi}{\mathbf{g}_{i}}

\newcommand{\hok}{\mathbf{h}_{1,k}}

\newcommand{\gok}{\mathbf{g}_{1,k}}
\newcommand{\gtk}{\mathbf{g}_{2,k}}
\newcommand{\goi}{\mathbf{g}_{1,i}}

\newcommand{\gohk}{\hat{\mathbf{g}}_{1,k}}
\newcommand{\gthk}{\hat{\mathbf{g}}_{2,k}}
\newcommand{\gohi}{\hat{\mathbf{g}}_{1,i}}

\newcommand{\gohm}{\hat{\mathbf{g}}_{1,m}}
\newcommand{\gthm}{\hat{\mathbf{g}}_{2,m}}

\newcommand{\etk}{\mathbf{e}_{2,k}}

\newcommand{\am}{\alpha_{\mathrm{m}}}
\newcommand{\az}{\alpha_{\mathrm{z}}}

\newcommand{\Gohi}{\big(\Goh^H\Goh\big)^{-1}}
\newcommand{\Gthi}{\big(\Gth^H\Gth\big)^{-1}}
\begin{document}
\title{Hybrid Processing Design for Multipair Massive MIMO Relaying with Channel Spatial Correlation}
\author{ Milad~Fozooni, Hien Quoc Ngo,~\IEEEmembership{Member,~IEEE,} Michail~Matthaiou,~\IEEEmembership{Senior Member,~IEEE,} Shi~Jin,~\IEEEmembership{Senior Member,~IEEE,} and George~C.~Alexandropoulos,~\IEEEmembership{Senior Member,~IEEE}  
\thanks{Manuscript received September 26, 2017; revised March 3, 2018 and June 9, 2018; accepted August 13, 2018. The associate editor coordinating the review of this paper and approving it for publication was  M. Cenk. Gursoy.}
\thanks{M. Fozooni, H. Q. Ngo and M. Matthaiou  are with the Institute of Electronics, Communications and Information Technology (ECIT), Queen's University Belfast, Belfast, BT3 9DT, U.K. (e-mail: \{mfozooni01, hien.ngo, m.matthaiou\}@qub.ac.uk).
\mbox{S. Jin} is with the National Mobile Communications Research Laboratory, Southeast University, Nanjing 210096, China (e-mail: jinshi@seu.edu.cn).
G. C. Alexandropoulos is with the Mathematical and Algorithmic Sciences Lab, Paris Research Center, Huawei Technologies France SASU, 92100 Boulogne-Billancourt, France (e-mail: george.alexandropoulos@huawei.com). 
The work of M. Matthaiou was supported by EPSRC, UK, under grant EP/P000673/1. The work of S. Jin was supported in part by the National Science Foundation (NSFC) for Distinguished Young Scholars of China with Grant 61625106.}}
\maketitle
\begin{abstract}
Massive multiple-input multiple-output (MIMO) avails of simple transceiver design which can tackle many drawbacks of relay systems in terms of complicated signal processing, latency, and noise amplification. However, the cost and circuit complexity of having one radio frequency (RF) chain dedicated to each antenna element are prohibitive in practice. In this paper, we address this critical issue in amplify-and-forward (AF) relay systems using a hybrid analog and digital (A/D) transceiver structure. More specifically, leveraging the channel long-term properties, we design the analog beamformer which aims to minimize the channel estimation error and remain invariant over a long timescale. Then, the beamforming is completed by simple digital signal processing, i.e., maximum ratio combining/maximum ratio transmission (MRC/MRT) or zero-forcing (ZF) in the baseband domain. We present analytical bounds on the achievable spectral efficiency taking into account the spatial correlation and imperfect channel state information at the relay station. Our analytical results reveal that the hybrid A/D structure with ZF digital processor exploits spatial correlation and offers a higher spectral efficiency compared to the hybrid A/D structure with MRC/MRT scheme. Our numerical results showcase that the hybrid A/D beamforming design captures nearly $95\%$ of the spectral efficiency of a fully digital AF relaying topology even by removing half of the RF chains. It is also shown that the hybrid A/D structure is robust to coarse quantization, and even with $2$-bit resolution, the system can achieve more than $93\%$ of the spectral efficiency  offered by the same hybrid A/D topology with infinite resolution phase shifters.
\end{abstract}
\section{Introduction}
\label{Introduction}
Cooperative relaying is a standard technique that improves the network performance due to its great ability in reducing transmit power and extending coverage, especially at the cell edges \cite{phan2009power,rong2009unified}. However, an excessive burden of signal processing is typically imposed upon the relay nodes. Thus, the complexity of relaying has always been an important issue as even a small delay in process means a loss of valuable physical resources, e.g., time, which should be available for other purposes, like backhaul and access link operations \cite{wang2016Low,hoymann2012relaying}. 
Recently, massive multiple-input multiple-output (MIMO) technology is becoming one of the most promising solutions to realize the challenging requirements of 5G wireless networks, such as massive connectivity, high-speed data transmissions, and simple signal processing. In massive MIMO, the channel vectors are (nearly) pairwisely orthogonal, and hence, linear processing schemes, such as maximum ratio combining/maximum ratio transmission (MRC/MRT), zero-forcing (ZF), and minimum mean-square error (MMSE) are nearly optimal. Not surprisingly, massive MIMO relaying has very recently received a great deal of research interest from different viewpoints. In \cite{suraweera2013multi}, the asymptotic performance of one-way massive relaying was analyzed in three different cases to scale down the radiated power by the number of active antennas at the relay station. The energy efficiency of two-way relaying with unlimited relay nodes was derived in \cite{ngo2013large}, while some other works in this context can be found in \cite{cui2014multi,kong2016multi,jin2015ergodic}. The potential of massive relaying to mitigate self-interference in full-duplex relaying was originally investigated in the seminal work of \cite{ngo2014multipair}. 

Generally, the practical deployment of massive MIMO relaying brings new practical \mbox{challenges}. More explicitly, having one radio frequency (RF) chain behind each antenna, in all above cases, will increase significantly the power consumption, digital signal processor (DSP) complexity, implementation/maintenance cost, and circuit complexity. Recently, this critical issue has been considered in cellular systems in which hybrid analog and digital (A/D) architectures, where the overall beamformer consists of a low dimensional digital baseband processor and an analog RF beamformer implemented by phase shifters, were proposed and analyzed \cite{lee2017hybrid,sohrabi2016hybrid,han2015large,el2014spatially,mendez2016hybrid,zhu2017secure,molisch2017hybrid}.
However, there is a dearth of literature considering the hybrid A/D solution for relaying systems \cite{lee2014af,fozooni2015massive,xu2017spectral}. The authors in \cite{lee2014af} proposed an algorithm based on the singular value decomposition (SVD) to maximize the average rate for millimeter wave MIMO systems. However, they did not derive any closed-form expression for the achievable rate. Moreover, the results are obtained under the idealistic assumptions of perfect channel state information (CSI) and accurate phase shifters. In \cite{fozooni2015massive}, the authors addressed part of these issues by relaxing the resolution of phase shifters to arbitrary quantization bits, and also deriving a closed-form expression for the achievable spectral efficiency. However, in both papers, phase shifters should be able to adapt to the quick variations of the propagation channels over time. This phase adaptation, not only requires perfect CSI at the relay station, but also is a challenging task due to the inherently smaller flexibility of analog beamformers compared to the digital ones. 

Altogether, there exist some formidable issues within the state-of-the-art of relaying with hybrid A/D processing. First, analog beamformers are usually designed to adapt to the quick variations of the propagation channels based on iterative algorithms. However, this methodology not only increases the complexity of signal processing, but also requires perfect CSI at the relay station \cite{fozooni2015massive,lee2014af}. Second, the algorithmic designs for hybrid A/D beamforming \cite{kim2015mse,adhikary2013joint,liu2014phase,tan2017spectral} fall short of providing an insightful closed-form expression for the most important performance metrics, e.g., achievable spectral efficiency. Third, it has been long recognized that the spectral efficiency of point-to-point MIMO is deteriorated due to spatial correlation \cite{tulino2005impact}. Nevertheless, spatial correlation in multiuser MIMO systems can be exploited in the transceiver design offering performance improvements \cite{alexandropoulos2015maximal,kim2008effects}. To the best of our knowledge, there is no prior work in the context of hybrid A/D MIMO relaying under spatially correlated fading channels.

In this paper, motivated by the above discussion we adjust the analog beamformers to the slow variations of channel statistics rather than the short-term fluctuations of channel. Thus, we will design our correlation-based analog beamformers to leverage the long-term channel spatial selectivity under the hybrid A/D structure relaying. Our main contributions are summarized as: 
\begin{itemize}
\item We consider a relay station with a hybrid A/D architecture, and take the spatial correlation and imperfect CSI into account. In this scenario, we explicitly evaluate the role of digital beamformer by developing analytical bounds on the spectral efficiency for the prevalent digital schemes, i.e., MRC/MRT and ZF. Our closed-form expressions, involve only the statistical parameters of the channels. Also, our numerical results reveal that the ZF scheme benefits of channel correlation and offers higher spectral efficiency compared to the MRC/MRT scheme, even for low signal-to-noise ratio (SNR) values. 

\item We design a correlation-based analog beamformer which exploits the long-term eigenmodes of the propagation channel to minimize the channel estimation error and, eventually, maximize the achievable spectral efficiency. Our analytical results showcase that, although, multiplexing and array gain are obviously restricted under this hybrid A/D architecture, the system can still avail of promising spatial diversity gain. As a matter of fact, our simulation results illustrate that with only $50$ RF chains, in an example massive MIMO system with $128$ antennas, the hybrid A/D structure can nearly capture $95 \%$ of the spectral efficiency achieved by the conventional fully digital structure with $128$ RF chains.      

\item We numerically evaluate the spectral efficiency of hybrid A/D structure under quantized phase shifters and covariance estimation error. We observe that the hybrid A/D configuration paradigm is robust to phase quantization, and this observation is more pronounced for hybrid A/D structure utilizing the MRC/MRT scheme.   
\end{itemize}
\noindent The rest of the paper is structured as: In Section \ref{System Model Journal}, we present the hybrid structure and channel estimation scheme. Section \ref{Spectral Efficiency Analysis Journal} investigates the system spectral efficiency under two prevalent digital beamforming schemes, while the design of the analog part is delineated in Section \ref{Analog Beamformer Design Journal}. Simulation results are presented in Section \ref{Simulation Results Journal} before concluding the paper in Section \ref{Conclusion Journal}.

\textit{Notation}: The boldface upper and lower case symbols denote matrices and vectors, respectively. The $\left(i,j\right)$-th element of a matrix $\mathbf{A}$ is denoted by $\left[\mathbf{A}\right]_{ij}$, and $\mathbf{a}_i$ refers to its $i$-th column. Moreover, the notations $\left(\cdot\right)^T$, $\left(\cdot\right)^*$, $\left(\cdot\right)^H$, $\mathrm{Tr}\left(\cdot\right)$, and $\|{\cdot}\|$ represent the transpose, conjugate, conjugate transpose, trace, and Frobenius norm (Euclidean norm for vectors), respectively. Also, $\mathbb{E}\left[\cdot\right]$ denotes the expectation operation, $\mathrm{Var}\left(\cdot\right)$ represents the variance of a random variable, while $\mathrm{Cov}\left(\cdot\right)$ denotes the covariance of a random vector. Additionally, $\lceil \cdot \rceil$ means rounding up to the nearest integer. The symbol $\mathrm{diag}\left\{\alpha_1,\ldots, \alpha_N\right\}$ denotes an $N \times N $ diagonal matrix with the vector $\left[\alpha_1,\ldots, \alpha_N \right]^T$ placed on the main diagonal. Also, $\mathcal{CN}\left(\mathbf{a},\mathbf{A}\right)$ shows a complex Gaussian random vector with mean $\mathbf{a}$ and covariance matrix $\mathbf{A}$. Finally, the symbols $\stackrel{\mathrm{a.s.}}{\longrightarrow}$ and $\stackrel{\mathrm{q.m.}}{\longrightarrow}$ signify the almost sure convergence and quartic mean limit, respectively.             
\section{System Model}
\label{System Model Journal}
\subsection{Signal Model and Hybrid Architecture}
\begin{figure*}[!t]
\centering
		\includegraphics[width=6in]{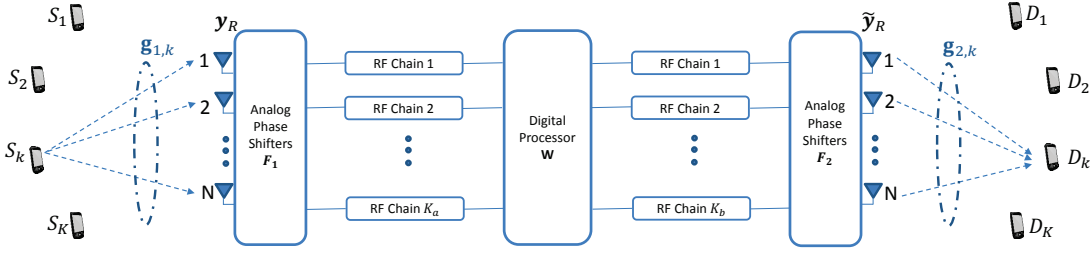}
		\caption{Block diagram of a multipair relay system with a baseband digital processor 
		combined with two analog RF beamformers.}
		\label{system_model_fig}
\end{figure*}

Consider a cooperative wireless system as depicted in Fig. \ref{system_model_fig}, where $K$ designated pairs of single antenna users, with one-to-one pairing $S_k$--$D_k$ with $k=1, \ldots, K$, intend to communicate with each other via a single half-duplex relay station equipped with very large number of service antennas $N$ on each side. We assume the most popular cooperation protocol, i.e., amplify-and-forward (AF) relaying, where the relay station simply amplifies the received signals from the sources, and send them to the destinations. In this protocol, the relay station serves all users in the same frequency band and by leveraging a time-division duplex (TDD) operation. We focus on a narrowband flat-fading propagation channel, while we note that the same results can be extended to wideband channels for orthogonal frequency division multiplexing (OFDM) systems, where the flat-fading assumption can hold for each frequency subcarrier \cite{molisch2017hybrid}.

Considering that users are randomly located around the relay station, we assume that the propagation channels at the users' terminals are uncorrelated due to their large distance from each other, whereas there exists correlation among the relay antennas on both sides of the relay station. This correlation  occurs if relay antennas are insufficiently spaced or there are a limited number of scatters surrounding the relay station. We also assume that there are no direct links between $K$ pairs due to the heavy shadowing and/or path loss attenuation. For the simplicity of analysis, and for obtaining a clear understanding of the impact of the antennas correlation, large-scale fading is neglected. Nevertheless, it is notable that our results can be readily extended to propagation channels which include large-scale fading.\footnote{The impact of large-scale fading can be compensated by a power control scheme at the relay station which is proportional to the inverse of the channel long-term attenuation, including path-loss and shadow fading \cite{bjornson2015optimal,bjornson2016massive}.}  Most importantly, we assume that CSI is not available at the users' nodes, and the relay station has to estimate the channels via uplink pilots during the training phase. 

{\color{black}To alleviate the hardware complexity and reduce the power consumption, we deploy $K_a+K_b$ transceiver RF chains at the relay station which are far smaller than the number of relay antennas $K \leq K_a , K_b \ll N$. During the pilot training, all RF chains work in the receive mode. However, during data payload, $K_a$ RF chains receive the signals and $K_b$ RF chains retransmit the received signals. These two groups of RF chains are connected to the relay antennas via an analog combiner $\Fo \in\mathbb{C}^{K_a\times N}$ and precoder $\Ft \in \mathbb{C}^{K_b\times N}$, respectively. For the ease of fabrication/implementation cost, these analog devices contain a bank of phase shifters \cite{mendez2016hybrid}.} 
Under this configuration setup, we mathematically model the received and transmitted signals at the relay station and also the received signals at $K$ destinations in the following way
\begin{align}
\hspace{-2pt}\mathbf{y}_R&=\sqrt{P_u}\mathbf{G}_1\mathbf{x}+\nr,\\
\hspace{-2pt}\tilde{\mathbf{y}}_R&=\mathbf{F}_2^H\mathbf{W}\mathbf{F}_1\mathbf{y}_R \label{Relay Transmit Data},\\
\hspace{-2pt}\mathbf{y}_D&=\sqrt{P_u}\Gt^H\Ft^H\W\Fo\Go\mathbf{x}+ \Gt^H\Ft^H\W\Fo\mathbf{n}_R+\mathbf{n}_D \label{received signal 1},
\end{align}
where $P_u$ denotes the average transmitted power of each source, and $\mathbf{x}=\left[x_1,x_2,\ldots, x_K\right]^T$ is the zero-mean information vector with $\mathbb{E}\left[\mathbf{x}\mathbf{x}^H\right]=\mathbf{I}_K$. Also, the $N$-dimensional vector $\mathbf{n}_R$ and $K$-dimensional vector $\mathbf{n}_D$ are additive white Gaussian noise (AWGN) vectors consisting of independent and identically distributed (i.i.d.) $\mathcal{CN}\left(0, \sigma_{n_R}^2\right)$ and $\mathcal{CN}\left(0, \sigma_{n_D}^2\right)$ random variables, respectively. The matrix $\W \in \mathbb{C}^{K_b \times K_a}$ is the digital processor which adjusts both the modulus and phase of the baseband received signal at the relay station. The propagation channels from the $K$ sources to the relay, and from the relay to the destinations are expressed as $\Go$, $\Gt \in \mathbb{C}^{N \times K}$, respectively. We represent the spatial covariance matrices between the relay antennas by $\Ro$, $\Rt \in \mathbb{C}^{N \times N}$ which have been already incorporated into the channel matrices, i.e., $\Go=\Ro^{\frac{1}{2}}\Ho$ and $\Gt=\Rt^{\frac{1}{2}}\Ht$. In this model, $\Ho$ and $\Ht \in \mathbb{C}^{N \times K}$ denote the small-scale fading with i.i.d. $\mathcal{CN}\left(0, 1\right)$ elements. Then, from (\ref{received signal 1}), the received signal at the $k$-destination  can be extracted as
\begin{align}
\label{received signal 2}
y_{D_k}\hspace{-3pt}=\hspace{-1pt}&\sqrt{P_u}\gtk^H\mathbf{F}_2^H\W\Fo\gok x_k\hspace{-2pt}+\hspace{-2pt}\sqrt{P_u}\sum_{i\neq k}^K \gtk^H\Ft^H\W\Fo\goi x_i \nonumber \\
&+\gtk^H\Ft^H\W\Fo\nr+n_{D_k},
\end{align}
\noindent where $n_{D_k}$ is AWGN at the $k$-th destination. In this aggregated received signal, the first term refers to the desired signal and the second term signifies inter-user interference, while the last two terms represent the compound noise.\\ 
\indent This system model is under two major practical constraints. First, the analog combiner and precoder just contribute to the phase alignment as they are usually implemented using a bank of analog phase shifters. In this light, we have to assign an equal modulus to all the entries of matrices $\Fo$ and $\Ft$. We fix this modulus by $1/\sqrt{N}$ to avoid an unlimited gain in the analog domain. 
Second, the AF relay station receives the signals from all sources, and boosts them up to a certain level of power $P_r$ before transmitting to the destinations. Thus, we consider the following long-term power constraint
\begin{align}
\label{constraint}
\mathbb{E}\Big[\mathrm{Tr}\big(\tilde{\mathbf{y}}_R\tilde{\mathbf{y}}_R^H\big)\Big]=P_r.
\end{align}
\subsection{Channel Estimation}
\label{Channel Estimation Method}
To capture the advantages that massive MIMO relay can offer, CSI is required at the relay station. This CSI is exploited to design the digital processor $\W$. The CSI acquisition at the relay station is done via the uplink pilots transmitted from the sources and the destinations. Let $\tc$ be the number of transmission symbols over each time-frequency coherence block, and $\tp$ be the number of symbols for uplink training. Then, the remaining part, $\tc-\tp$, is used for downlink data transmission. We assume that pilot sequences sent from all sources and destinations are mutually orthogonal. This requires $2K\leq \tp\leq \tc$.
Let us stack all of these pilot sequences into the matrices $\sqrt{\tp P_p}\Phio \in \mathbb{C}^{\tp \times K}$ and $\sqrt{\tp P_p}\Phit \in \mathbb{C}^{\tp \times K}$, where $P_p$ denotes the average pilot power. Then, we have $\Phio^H\Phio=\mathbf{I}_K$, $\Phit^H\Phit=\mathbf{I}_K$ and $\Phio^H\Phit=\mathbf{0}_K$. The baseband received pilot signals at the receive and transmit sides of the relay station before the DSP unit can be expressed, respectively, as\footnote{Note that the process for the channel estimation should be done in the digital domain. Thus, we take into account the signals after $\Fo$, $\Ft$ which exactly refer to the baseband signals before the DSP unit.}

\begin{align}
\Ypr&=\sqrt{\tp P_p}\Fo\Go\Phio^T+ \Npr,\\
\Ypt&=\sqrt{\tp P_p}\Ft\Gt\Phit^T+ \Npt,
\end{align} 
where $\Npr \in \mathbb{C}^{K_a \times \tp}$ and $\Npt \in \mathbb{C}^{K_b \times \tp}$ are AWGN matrices including i.i.d. $\mathcal{CN}\left(0,1\right)$ entries. Hereafter, we just derive the results for the source side and the same result can be similarly deduced for the destination side. We can obtain the matrix $\Yo$ after projecting $\Ypr$ onto \mbox{$\Phio$}   
\begin{align}
\label{pilot}
 \Yo=\Ypr\Phio^*=\sqrt{\tp P_p}\Fo\Ro^{\frac{1}{2}}\Ho+\No,
\end{align}
\noindent where the entries of $\No \in \mathbb{C}^{K_a \times K}$ follow the same distribution as the entries of $\Npr$ due to the fact that the columns of $\Phio$ are orthonormal. Then, for a given channel covariance matrix $\Ro$, the MMSE estimate of $\Go$ can be obtained by
\begin{align}
\label{Estimation}
\Goh&=\sqrt{\tp P_p}\Ro\Fo^H\Big(\tp P_p \Fo\Ro\Fo^H + \mathbf{I}_{K_a}\Big)^{-1} \Yo.
\end{align}
Since the MMSE estimator is a linear transformation, we can conclude that the channel estimation matrices $\Goh$, $\Gth$ and their errors $\Eo$, $\Et$ are Gaussian random matrices, such that
\begin{align}
\Go&=\Goh+\Eo \label{estimation error1}, \\
\Gt&=\Gth+\Et. \label{estimation error2}
\end{align}
\indent Although the MMSE estimator has been developed based on the observation matrix $\Yo$ in (\ref{Estimation}), we now reformulate this estimator based on an auxiliary Gaussian random matrix $\Hoh \in \mathbb{C}^{N \times K}$ with i.i.d. $\mathcal{CN}\left(0, 1\right)$ entries to simplify our analysis in the subsequent sections. We indicate that both $\Goh$ and $\Hoh$ depend on the observation matrix $\Yo$ and, consequently, depend on the propagation channel $\Go$. It is notable that the following equality in distribution for $\Goh$ holds due to the same mean value and same second-order expectations for both sides:    
\begin{align}
\Goh&=\Bigg(\frac{\mathbb{E}\big[\Goh\Goh^H\big]}{\mathrm{Tr}\Big(\mathbb{E}\big[\Goh^H\Goh\big]\Big)}\Bigg)^{\frac{1}{2}} \Hoh \Big(\mathbb{E}\big[\Goh^H\Goh\big]\Big)^{\frac{1}{2}},
\end{align}
\noindent The latter expression follows from Lemma \ref{Lemma1_MRC} in Appendix \ref{Prerequisite Lemmas}. By leveraging this property of Gaussian random matrices, the channel estimate can be reformulated as a product of matrices which suits to our subsequent analysis 
\begin{align}
\hspace{-4pt}\Goh&\hspace{-3pt}=\hspace{-3pt}\bigg(\hspace{-2pt}\tp P_p \Ro \Fo^H \Big(\hspace{-2pt}\tp P_p \Fo\Ro\Fo^H \hspace{-2pt}+ \hspace{-2pt} \mathbf{I}_{K_a}\hspace{-2pt}\Big)^{-1}\hspace{-2pt} \Fo \Ro \bigg)^{\frac{1}{2}}\hspace{-2pt}\Hoh, \label{former}\\
&=\bigg(\Ro-\Big(\Ro^{-1}+\tp P_p \Fo^H \Fo\Big)^{-1}\bigg)^{\frac{1}{2}}\Hoh.
\end{align}
The latter equality is derived by invoking the matrix inversion lemma if $\Ro$ is invertible; otherwise, we can proceed with the former equation (\ref{former}) without loss of generality. We can derive the same result in a similar fashion for the error matrix. we now gather both results into the following simple expressions
\begin{align}
&\Goh=\Uo^{\frac{1}{2}}\Hoh \label{Channel1},\\
&\Eo=\Ueo^{\frac{1}{2}}\Heo \label{Error1},
\end{align}
where $\Heo \in \mathbb{C}^{N \times K}$ is another auxiliary random matrix independent of $\Hoh$, and its entries are independent $\mathcal{CN}\left(0,1\right)$. Furthermore, $\Ueo$ and $\Uo$ are the error covariance matrix and estimation covariance matrix, given respectively by
\begin{align}
&\Ueo\stackrel{\Delta}{=}\Big(\Ro^{-1}+\tp P_p \Fo^H \Fo\Big)^{-1}, \\
&\Uo\stackrel{\Delta}{=}\Ro-\Ueo.
\end{align}
\begin{figure*}[th]
\begin{align}
\label{Hassibi_MRC}
y_{D_k}&=\underbrace{\mathbb{E}\Big[\sqrt{P_u}  \gthk^H \Ft^H\W\Fo \gohk\Big] x_k}_{\mathrm{desired \, \, signal}}+\underbrace{\sqrt{P_u} \gthk^H \Ft^H\W\Fo \gohk x_k-\mathbb{E}\Big[\sqrt{P_u} \gthk^H \Ft^H\W\Fo \gohk \Big]x_k}_{\mathrm{fluctuation\,\, of \,\, desired \, \, signal}} \nonumber \\
&+\underbrace{\sqrt{P_u}\sum_{i\neq k}^K \gthk^H\Ft^H\W\Fo\gohi x_i}_{\mathrm{inter-user \, \, interference}} + \underbrace{\sqrt{P_u}  \gthk^H \Ft^H\W\Fo \Eo \mathbf{x} }_{\mathrm{estimation \, \, error}}+ \underbrace{\sqrt{P_u}  \etk^H \Ft^H\W\Fo \Goh \mathbf{x}}_{\mathrm{estimation\, \, error}} \nonumber \\
&+\underbrace{\sqrt{P_u} \etk^H \Ft^H\W\Fo \Eo \mathbf{x}}_{\mathrm{estimation \, \, error}}+ \underbrace{\gthk^H\Ft^H\W\Fo \nr+\etk^H\Ft^H\W\Fo \nr+n_{D_k}}_{\mathrm{aggregated \,\, noise\,\, \& \,\, estimation \, \, error}}. 
\end{align}
\hrule
\end{figure*}
\section{Spectral Efficiency Analysis}
\label{Spectral Efficiency Analysis Journal}
In this section, we analytically evaluate the spectral efficiency of hybrid A/D configuration paradigm to get better insights into how analog beamformers affect the system performance. The spectral efficiency is derived via the ``use and forget" technique which is commonly used in the context of massive MIMO \cite{marzetta2016fundamentals}. Let us recall (\ref{received signal 2}) and take the channel estimation matrices (\ref{estimation error1}), (\ref{estimation error2}) into account, then we have the signal model presented in (\ref{Hassibi_MRC}) at the top of next page.  
\noindent In this model, we identify the first term as the desirable signal since only the channel estimates available at the relay station. However, it is reasonable to assume that the long-term statistics of the channels are available at the destinations. Then, we obtain an achievable sum spectral efficiency as 
\begin{align}
R=&\,\frac{\tc-\tp}{2\tc}\sum_{k=1}^K\log_2\Big(1+\mathrm{SINR}_k\Big), \label{Rate} 
\end{align}
\noindent where,
\begin{align}
\hspace{-10pt}\mathrm{SINR}_k=&\frac{P_ut_0}{P_u \left(t_1+t_2+t_3+t_4+t_5\right)+  \left(t_6+t_7\right)+ \sigma_{n_D}^2}\label{SINR}.
\end{align}
In (\ref{Rate}), the factor $\frac{\tc-\tp}{2\tc}$ denotes the penalty loss due to the half-duplex relaying operation and the channel estimation overhead. In (\ref{SINR}), $t_0, t_1, \ldots, t_7$ are defined as 
$t_0 =\bigg|\mathbb{E}\Big[\gthk^H\Ft^H\W\Fo\gohk\Big]\bigg|^2$, 
$t_1= \mathrm{\mathrm{Cov}}\Big(\gthk^H\Ft^H\W\Fo \gohk \Big)$,
$t_2=\mathbb{E}\bigg[\Big|\sum_{i\neq k}\gthk^H \Ft^H\W\Fo \gohi x_i\Big|^2\bigg]$,
$t_3=\mathbb{E}\bigg[\Big|\gthk^H \Ft^H\W\Fo \Eo\mathbf{x}\Big|^2\bigg]$, 
$t_4=\mathbb{E}\bigg[\Big|\etk^H \Ft^H\W\Fo \Goh \mathbf{x}\Big|^2\bigg]$,
$t_5=\mathbb{E}\bigg[\Big|\etk^H \Ft^H\W\Fo \Eo \mathbf{x}\Big|^2\bigg]$,
$t_6=\mathbb{E}\bigg[\Big|\gthk^H \Ft^H\W\Fo \tilde{\mathbf{n}}_R\Big|^2\bigg]$, and
$t_7=\mathbb{E}\bigg[\Big|\etk^H\Ft^H\W\Fo \tilde{\mathbf{n}}_R \Big|^2\bigg]$,
where for the sake of simplicity we omitted index $k$ that indicate these terms that correspond to the $k$-th user pair. It is also notable that in the above equations $t_0 - t_7$, the expectation is over all the random variables: information symbols, estimated channels, channels estimation errors, and AWGN. Since finding the optimal $\W$ is a demanding task due to the non-convex nature of the problem, in the subsections that follow we adopt MRC/MRT and ZF digital processing to simplify the spectral efficiency expressions under these simple but asymptotically robust linear schemes. Then, given the closed-form expression in this section, we can design our analog beamformer in the next section (Section \ref{Analog Beamformer Design Journal}) to minimize the channel estimation error and, consequently, improve the spectral efficiency.   
\subsection{MRC/MRT Digital Processor}
\label{MRC Scheme Journal}
MRC/MRT is a simple scheme that coherently combines the received signals, and then sends them toward the destinations. With MRC/MRT, $\mathbf{W}$ is chosen so that $\Ft^H\W\Fo=\am \Gth\Goh^H$, where $\am$ is a relay amplification factor chosen to satisfy the long-term constraint in (\ref{constraint}). Hereby, we certify that the aforementioned equation has the following solution for \textbf{W}:
\begin{align} 
{\color{black}{\W=\am \left(\Ft\Ft^H\right)^{-1}\Ft\Gth\Goh^H\Fo^H\left(\Fo\Fo^H\right)^{-1}}}. \label{mrc_processor}
\end{align} 
Next, we provide some propositions that further simplify the SINR terms $t_0 - t_7$, and provide a closed-form expression for the spectral efficiency which only involves the long-term statistics of the propagation channels. For ease of exposition, we use superscript ``mrc'' to denote the MRC/MRT scheme.   
\begin{Term1_MRC}
\label{Term1_MRC}
With MRC/MRT, the mathematical terms corresponding to the desired signal and its fluctuations, $t_0$ and $t_1$, can be expressed in closed-form, respectively, as
\begin{align}
t_0^{\mathrm{mrc}} \,\,=\,\, & \am^2 \mathrm{Tr}^2\big(\Uo\big) \mathrm{Tr}^2\big(\Ut\big),\\
t_1^{\mathrm{mrc}} \,\,=\,\, & \am^2\bigg(\mathrm{Tr}^2\big(\Uo\big)\big\|\Ut\big\|^2 \nonumber \\
&+ \mathrm{Tr}^2\big(\Ut\big)\big\|\Uo\big\|^2  + K\big\|\Uo\big\|^2\big\|\Ut\big\|^2 \bigg).   \label{Fluctuation Term} 
\end{align}
\end{Term1_MRC}
\begin{proof}
The proof is relegated in Appendix \ref{Term1_MRC_ap}.
\end{proof}
\begin{Term2_MRC}
\label{Term2_MRC}
With MRC/MRT, the mathematical term corresponding to inter-user interference, $t_2$, can be expressed in closed-form as
\begin{align}
t_2^{\mathrm{mrc}}=&\am^2(K-1)\Big(\mathrm{Tr}^2\big(\Uo\big)\big\|\Ut\big\|^2 \nonumber \\ & + \mathrm{Tr}^2\big(\Ut\big)\big\|\Uo\big\|^2 + K\big\|\Uo\big\|^2\big\|\Ut\big\|^2 \Big).   \label{Intereference_MRC}    
\end{align}
\end{Term2_MRC}
\begin{proof}
\mbox{Follows the methodology of Proposition \ref {Term1_MRC}.}
\end{proof}
\begin{Term3_MRC}
\label{Term3_MRC}
With MRC/MRT, the mathematical term $t_3$ is given by
\begin{align}
t_3^{\mathrm{mrc}} = K \am^2 \mathrm{Tr}\big(\Uo\Ueo\big) \Big(\mathrm{Tr}^2\left(\Ut\right) + K\left\|\Ut\right\|^2\Big). 
\end{align}
\end{Term3_MRC}
\begin{proof}
See Appendix \ref{Term3_MRC_ap}.
\end{proof}	
\begin{Term4_MRC}
\label{Term4_MRC}
With MRC/MRT, the mathematical term $t_4$ can be expressed in closed-form as
\begin{align}
t_4^{\mathrm{mrc}}= K \am^2 \mathrm{Tr}\left( \Ut\Uet \right) \Big(\mathrm{Tr}^2\left(\Uo\right) + K\left\|\Uo\right\|^2\Big).
\end{align}
\end{Term4_MRC}
\begin{proof}
Appendix \ref{Term4_MRC_ap} includes the proof.
\end{proof}	
\noindent A similar methodology can be applied for the rest of terms $t_5 - t_7$ to finally obtain the following results
\begin{align}
t_5^{\mathrm{mrc}} = \, & K^2 \am^2 \mathrm{Tr}\left( \Uo\Ueo \right)\mathrm{Tr}\left( \Ut\Uet \right), \\
t_6^{\mathrm{mrc}} = \, & \am^2 \sigma_{n_R}^2 \mathrm{Tr}\left( \Uo\right)\Big(\mathrm{Tr}^2\left( \Ut\right)+K \big\|\Ut\big\|^2\Big),\\
t_7^{\mathrm{mrc}} = \, & \am^2 \sigma_{n_R}^2 K \mathrm{Tr}\left( \Uo\right)\mathrm{Tr}\left( \Ut\Uet \right).
\end{align} 
The power amplification factor $\am$ is enforced by the long-term power constraint at the AF relay station. Hence, starting from (\ref{constraint}), recalling the relay transformation matrix in (\ref{Relay Transmit Data}), and then proceeding with the same strategy that we used to simplify the SINR terms, this amplification gain can be calculated by (\ref{alpha}) at the top of next page.
\begin{figure*}[t!]
\begin{align}
\label{alpha}
\am^2=\frac{P_r}{P_u K\mathrm{Tr}\left(\Ut \right)\!\Big(\mathrm{Tr}^2\left(\Uo\right) + K\left\|\Uo\right\|^2\Big)+ P_u K^2\mathrm{Tr}\left(\Uo \Ueo \right) \mathrm{Tr}\left(\Ut\right) +\sigma_{n_R}^2 K \mathrm{Tr}\left(\Uo\right)\mathrm{Tr}\left(\Ut\right)}.
\end{align}
\hrule
\end{figure*}
\subsection{ZF Digital Processor} 
\label{ZF Scheme Journal}
As previously shown in (\ref{Intereference_MRC}), the MRC/MRT scheme suffers a relatively high interference compared to other components: estimation error and AWGN, especially at large $K$. Motivated by this observation, we employ a ZF scheme at the DSP unit, and then derive an approximation of the spectral efficiency. The ZF receiver can be mathematically expressed based on (\ref{received signal 1}) so that $\Ft^H\W\Fo=\az \Gth\Gthi\Gohi\Goh^H$, where $\az$ is the relay amplification factor in ZF DSP. We note that there exists a solution for the above equation which is given as follows
\begin{align}
\W=\,\,&\az \left(\Ft\Ft^H\right)^{-1}\Ft \Gth\Gthi \nonumber \\ 
&\Gohi\Goh^H \Fo^H\left(\Fo\Fo^H\right)^{-1}. \label{zf_processor} 
\end{align}
\noindent For the ease of exposition, we use the superscript ``zf'' to denote the expressions related with the ZF digital processor. Considering the ZF strategy we have
\begin{align}
\hspace{-8pt} t_0^{\mathrm{zf}}=&\az^2, \,\,\,\,\,\,\	t_1^{\mathrm{zf}}= 0 , \,\,\,\,\,\,\ t_2^{\mathrm{zf}}= 0,\\
\hspace{-8pt} t_3^{\mathrm{zf}}=&\az^2 \bigg[\mathrm{Cov}\bigg(\Gohi\Goh^H\Eo\mathbf{x}\bigg)\bigg]_{\hspace{-2pt}k,k}\hspace{-2pt},\\
\hspace{-8pt} t_4^{\mathrm{zf}}=&\az^2 \bigg[\mathrm{Cov}\bigg(\Et^H\Gth\Gthi\mathbf{x}\bigg)\bigg]_{\hspace{-2pt}k,k}\hspace{-2pt},\\
\hspace{-8pt} t_5^{\mathrm{zf}}=&\hspace{-2pt}\az^2 \hspace{-0pt} \bigg[\hspace{-1pt}\mathrm{Cov}\hspace{-2pt}\bigg(\hspace{-2pt}\Et^H 
\Gth \Gthi \hspace{-2pt} \Gohi \hspace{-2pt} \Goh^H \Eo \mathbf{x}\hspace{-2pt}\bigg)\hspace{-2pt}\bigg]_{\hspace{-2pt}k,k}\hspace{-3pt},
\end{align}
\begin{align}
\hspace{-8pt} t_6^{\mathrm{zf}}=&\az^2 \bigg[\mathrm{Cov}\bigg(\Gohi \Goh^H \mathbf{n}_R\bigg)\bigg]_{\hspace{-2pt}k,k}\hspace{-2pt},\\
\hspace{-8pt} t_7^{\mathrm{zf}}=&\az^2 \bigg[\mathrm{Cov}\hspace{-2pt}\bigg(\hspace{-2pt}\Et^H \Gth \Gthi \hspace{-2pt}\Gohi \hspace{-2pt}\Goh^H \mathbf{n}_R\hspace{-2pt}\bigg)\hspace{-2pt}\bigg]_{\hspace{-2pt}k,k}\hspace{-2pt},
\end{align}
which can be approximated in the following manner.
\begin{Term1_ZF}
\label{Term1_ZF}
With ZF, the mathematical expression $t_3$ can be approximated by
\begin{align}
t_3^{\mathrm{zf}}\stackrel{\mathrm{a.s.}}{\longrightarrow}\frac{K \az^2\mathrm{Tr}\big(\Uo\Ueo\big)}{\mathrm{Tr}^2\big(\Uo\big)}.\end{align}
\end{Term1_ZF}
\begin{proof} 
See Appendix \ref{Term1_ZF_ap}.
\end{proof}
\noindent The other terms can be obtained in a similar spirit, though for the sake of brevity, we omit their proofs and just point out to their final results:
\begin{align}
t_4^{\mathrm{zf}}&\stackrel{\mathrm{a.s.}}{\longrightarrow} \frac{K\ \az^2\mathrm{Tr}\big(\Ut\Uet\big)}{\mathrm{Tr}^2\big(\Ut\big)},\\
t_5^{\mathrm{zf}}&\stackrel{\mathrm{a.s.}}{\longrightarrow} \frac{K^2 \az^2\mathrm{Tr}\big(\Uo\Ueo\big)\mathrm{Tr}\big(\Ut\Uet\big)}{\mathrm{Tr}^2\big(\Uo\big)\mathrm{Tr}^2\big(\Ut\big)},\nonumber \\
t_6^{\mathrm{zf}}&\stackrel{\mathrm{a.s.}}{\longrightarrow} \frac{\az^2\sigma_{n_R}^2}{\mathrm{Tr}\big(\Uo\big)},\\
t_7^{\mathrm{zf}}&\stackrel{\mathrm{a.s.}}{\longrightarrow} \frac{K\az^2\sigma_{n_R}^2\mathrm{Tr}\big(\Ut\Uet\big)}{\mathrm{Tr}\big(\Uo\big)\mathrm{Tr}^2\big(\Ut\big)},\\
\az^2&= \frac{P_r\mathrm{Tr}\big(\Uo\big)\mathrm{Tr}\big(\Ut\big)}{P_u K \mathrm{Tr}\big(\Uo\big)+\frac{P_u K^2\mathrm{Tr}\big( \Uo\Ueo\big)}{\mathrm{Tr}\big(\Uo\big)} + K \sigma_{n_R}^2}.\label{az}
\end{align} 
\section{Analog Beamformer Design}
\label{Analog Beamformer Design Journal}
\subsection{Analog Beamformer Design}
Massive MIMO can significantly increase the spectral efficiency thanks to its ability to provide a large multiplexing gain. However, this ability is somewhat restricted in highly correlated channels or in networks with few number of active users; in these cases, exploiting the multiplexing gain is not the main concern, and consequently a hybrid A/D structure with reduced number of RF chains becomes very relevant. However, a limited number of RF chains decreases the control of the DSP unit over the antenna arrays to finely steer the beams and place the nulls in predefined directions especially when perfect CSI is not available at the relay station. Although power and multiplexing gain are restricted in hybrid A/D structures, we will show that this topology can still deliver a reasonable and reliable spectral efficiency by extracting the best eigenmodes of the channel covariance matrix.
Therefore, in this section we design an analog beamformer based on the statistics of the propagation channel. 

\noindent \textit{Remark $1$}: Following the signal model in (\ref{received signal 1}) the analog beamformers, i.e., $\Fo$ and $\Ft$, are only involved in the channel estimates $\Goh$ and $\Gth$. Precisely speaking, by plugging (\ref{mrc_processor}) into (\ref{received signal 1}) the baseband received signal can be reformulated as 
\begin{align}
\hspace{-5pt} \mathbf{y}_D=\sqrt{P_u} \am \Gt^H\Gth \Goh^H\Go\mathbf{x}+ \Gt^H\Gt\Go^H\mathbf{n}_R+\mathbf{n}_D, \label{received_signal_2}
\end{align}
where we note that the analog beamformers $\Fo$ and $\Ft$ do not directly appear in the signal model, although, they implicitly contribute into the channel estimates $\Goh$ and $\Gth$. This technique confines the role of the analog beamformers into the channel estimates. The new signal model in (\ref{received_signal_2}) implies that the better estimate of the channel we gain, the higher spectral efficiency the system achieves.\footnote{ If we design $\W$ based on the low-dimensional effective channel  $\Fo\Goh$ and $\Ft\Gth$, the performance of the system will not be changed, but $\Fo$ and $\Ft$ will appear not only in the channel estimates $\Goh$ and $\Gth$, but also inside the signal model (\ref{received_signal_2}). Therefore, this methodology falls short of confining the role of $\Fo$ and $\Ft$ only into the channel estimates, and in this sense, the spectral efficiency analysis becomes quite challenging, if not impossible.} This discussion can be deduced for the ZF scheme in a same flavor. 

\begin{figure}[t!]
\centering
\hspace{-1.2cm}   \includegraphics[width=3.5in]{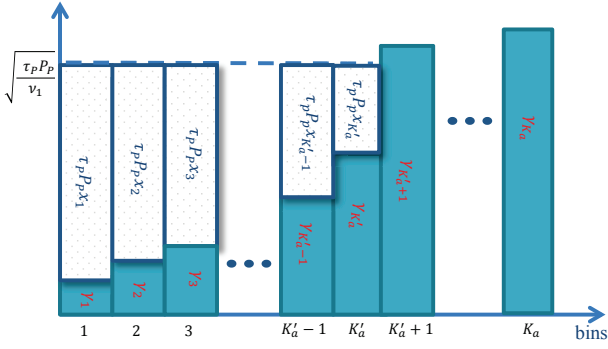}
  \vspace{-0.2cm}
	\caption{Water-filling structure.}
	\vspace{-0.3cm}
  \label{Water_Filling}
\end{figure}

Motivated by Remark $1$, we now design an analog beamformer in order to reduce the estimation errors, which in turn boosts the desired signal power, and finally improves the spectral efficiency. For simplicity, we confine our focus on designing the matrix $\Fo$, while the same results can be derived for $\Ft$. From (\ref{Error1}), the total estimation error is given by 
\begin{align}
\label{Estimation Error}
\varepsilon_1 =\mathbb{E}\bigg[\mathrm{Tr}\Big(\Eo\Eo^H\Big)\bigg]=K\mathrm{Tr}\big(\Ueo\big). 
\end{align}
Taking the practical constraints of analog beamformer into account, we can design the analog beamformer via the following optimization problem
\begin{align}
\label{Analog Optimization}
\min_{\Fo} \, \, \, &\varepsilon_1 \nonumber \\ 
\mathrm{s.t.}\, \, \, &\mathrm{Tr}\big(\Fo \Fo^H \big)\leq K_a, \nonumber \\ 
&\Fo \in \mathbb{C}^{K_a \times N}, \nonumber \\
&\Fo \in \mathcal{F},
\end{align}
where $\mathcal{F}$ denotes the set of $K_a \times N$ complex matrices with equal modulus. Since $\mathcal{F}$ is not a convex set, we ignore this constraint at this stage, but we will show the impact of this assumption in simulation results. Under this relaxation, we can rewrite the optimization problem as    
\begin{align}
\label{Digital Optimization}
\min_{\Fo} \, \, \, &\mathrm{Tr}\bigg(\Big(\Ro^{-1}+\tp P_p \Fo^H \Fo\Big)^{-1}\bigg) \nonumber\\ 
\mathrm{s.t.}\, \, \, &\mathrm{Tr}\big(\Fo \Fo^H \big)\leq K_a, \nonumber \\ 
&\Fo \in \mathbb{C}^{K_a \times N}. 
\end{align} 
Let  $\Fo=\Ufo\Sfo\Vfo^H$ and $\Ro=\Uro\Sro\Uro^H$ denote the singular value decomposition and eigen value decomposition of matrices $\Fo$ and $\Ro$, respectively. The minimum value of the objective function in (\ref{Digital Optimization}) can be achieved if the eigenvectors of $\Fo^H \Fo$ are chosen along with the eigenmodes of $\Ro^{-1}$, i.e., $\Vfo=\Uro$. By invoking this property, the optimization problem in hand can be further simplified to 
\begin{align}
\min_{\Sfo} \, \, \, &\mathrm{Tr}\bigg(\Big(\Sro^{-1}+\tp P_p \Sfo^H\Sfo\Big)^{-1}\bigg) \\ 
\mathrm{s.t.}\, \, \, &\mathrm{Tr}\big(\Sfo^H \Sfo \big)\leq K_a, \nonumber \\ 
&\Sfo \in \mathbb{C}^{K_a \times N}. \nonumber
\end{align} 
Let $x_i$ and $\gamma_i$ respectively denote the $i$-th biggest eigenvalue of $\Fo^H\Fo$ and the $i$-th smallest eigenvalue of $\Ro^{-1}$. Then, this problem can be reduced to a typical water-filling optimization as
\begin{align}
\min_{x_i} \, \, \, &\sum_{i=1}^{N}\bigg(\frac{1}{\gamma_i+\tp P_p x_i}\bigg) \nonumber \\ 
\mathrm{s.t.}\, \, \, &\sum_{i=1}^{N}x_i\leq K_a, \nonumber \\
& x_i\geq 0, \, \, \, \mathrm{for} \,\, i=1,\ldots,K_a, \nonumber\\
& x_i=0, \, \, \, \mathrm{for} \,\, i=K_a+1,\ldots,N. 
\end{align} 
By utilizing the Lagrangian duality, and considering the Karush--Kuhn--Tucker (KKT) condition for optimality, we deduce that the first $K'_a$ bins should be filled up to a certain level as illustrated in Fig. \ref{Water_Filling} so that  
\begin{align}
\sqrt{\frac{\tp P_p}{\nu_1}}=\gamma_i+\tp P_p x_i \label{Nu},
\end{align}
where $\nu_1$ is a constant number to satisfy the constraint $\sum_{i=1}^{N}x_i\leq K_a$ with equality. Then, after some simple mathematical manipulations we can find that 
\begin{align} 
 &\sqrt{\frac{\tp P_p}{\nu_1}}=\frac{\tp P_p K_a + \sum_{i=1}^{K^\prime_a}\gamma_i}{K^\prime_a}, \\
&x_i=\frac{1}{\tp P_p}\bigg(\frac{\tp P_p K_a + \sum_{i=1}^{K^\prime_a}{\gamma_i}}{K^\prime_a}-\gamma_i\bigg), \hspace{0.3cm} \mathrm{for} \,\, i=1,\ldots,K^\prime_a, \nonumber \\
&x_i=0, \hspace{5.0cm} \mathrm{for} \,\, i=K^\prime_a, \ldots, N. \nonumber
\end{align} 

It is noteworthy that the maximum rank of matrix $\Fo$ is $K_a$, hence the water-filling algorithm is applied to the first $K_a$ bins, whereas $K^\prime_a \leq K_a$ bins can be filled by this algorithm as illustrated in Fig. \ref{Water_Filling}. It is notable that the design of matrices $\Fo$ and $\Ft$ only depends on the long-term statistics of the channel according to the above discussion. The final design of the analog beamformer can be summarized as
\begin{align}
\Fo=\Ufo\Sfo\Uro^H,
\end{align} 
where $\Sfo$ is a $K_a \times N$ rectangular diagonal matrix with the vector $\left[\sqrt{x_1},\sqrt{x_2},\ldots,\sqrt{x_{K^\prime_a}},0,\ldots,0\right]^T$ on its main diagonal. The matrix $\Ufo$ does not play any role in the optimization problem (\ref{Digital Optimization}), hence any unitary matrix can be chosen. For convenience in our subsequent developments, we just assume $\Ufo=\mathbf{I}_N$ which leads to\footnote{Precisely speaking, $\Ufo$ can affect the original optimization problem (\ref{Analog Optimization}) where we need to design $\Ufo$ so that the matrix $\Fo$ falls within the set $\mathcal{F}$. It can also contribute to the system robustness, however, Monte-Carlo simulations will confirm that the results are still good enough by considering $\Ufo=\mathbf{I}_N$.} 
\begin{align}
\Fo=\Sfo\Uro^H.
\end{align}

\subsection{Discussion}
\label{Discussion}
Altogether, we could find a closed-form lower bound for the spectral efficiency for the MRC/MRT scheme, and also an approximation of the  spectral efficiency for the ZF scheme which depend on the long-term components of the channels as reflected by $\mathrm{Tr}\big(\mathbf{U}_i$\big), $\mathrm{Tr}\big(\mathbf{U}_i\mathbf{U}_{e_i}\big)$, and $\big\|\mathbf{U}_i\big\|$ for $i=1,2$. Proposition \ref{Trace Connection} will reduce the number of these components by finding a direct connection between these factors, and Proposition \ref{MMSE_error} will investigate the performance of the MMSE estimator for highly correlated channels under the proposed analog design.      
\begin{Trace Connection}
\label{Trace Connection}
It can be shown that $\mathrm{Tr}\big(\mathbf{U}_i$\big) and $\mathrm{Tr}\big(\mathbf{U}_i\mathbf{U}_{e_i}\big)$ for $i=1,2$ are directly connected to each other by the level of water-filling ceiling 
\begin{align}
\label{Relation}
\mathrm{Tr}\big(\mathbf{U}_i\big)=\sqrt{\frac{\tp P_p}{\nu_i}}\mathrm{Tr}\big(\mathbf{U}_i\mathbf{U}_{e_i}\big),  \hspace{0.5cm} \mathrm{for} \, \, i=1,2.
\end{align}    
\end{Trace Connection}
\begin{proof}
See Appendix \ref{Projection}.
\end{proof}
\begin{MMSE_error}
\label{MMSE_error}
Given the proposed analog beamformers $\Fo$ and $\Ft$, channels with higher spatial correlation result in a lower estimation error.
\end{MMSE_error}
\begin{proof}
\vspace{-0.1cm}
See Appendix \ref{MMSE_error_ap}.
\end{proof}
It is remarkable that the impact of $\mathrm{Tr}\big(\mathbf{U}_i\big)$ for $i=1,2$ is more pronounced in the numerator of the SINR in (\ref{SINR}), and hence, it can be treated as a desirable value that boosts the spectral efficiency. In contrast, $\mathrm{Tr}\big(\mathbf{U}_i\mathbf{U}_{e_i}\big)$, which can be interpreted as the inner product between the error covariance matrix $\mathbf{U}_{e_i}$ and estimation covariance matrix $\mathbf{U}_i$, deteriorates the spectral efficiency. Hence, Proposition \ref{Trace Connection} is seemingly useful to both mathematically and intuitively explain the relation between these two factors. This proposition numerically explains that for a given MMSE channel estimate, the impact of error estimation can be harnessed by a scalar, i.e., $\sqrt{\frac{\tp P_p}{\nu_i}}$ for $i=1,2$, which basically depends on the number of RF chains, pilot power, and the duration of pilot sequences. This is, of course, in line with the canonical concept of the pilot-based channel estimation originally observed in \cite{ngo2014massive,hassibi2003much}.

To obtain a clear understanding of how the different parameters affect the spectral efficiency in our analog beamformer design followed by the ZF DSP, we consider a rational case with the same channel covariance matrix and the same amount of RF chains at both sides of the relay station, i.e. $\Ro=\Rt$ and $K_a=K_b$. Also, we note that $t_5^{\mathrm{zf}}$, $t_7^{\mathrm{zf}}$ and a part of $\az$ are negligible compared to other terms.\footnote{By invoking Proposition \ref{Trace Connection} in (\ref{az}), we can readily conclude that the last two terms of denominator of $\az$ are much smaller than its first term.} Hence, we can derive a new simplified approximation as follows
\begin{align}
\bar{R}^\mathrm{zf}=\frac{K\left(\tc-\tp\right)}{2\tc} \log_2\left({1+\frac{\mathrm{Tr}\left(\U\right)}{2K\zeta+\frac{\sigma_{n_R}^2}{P_u}+ \frac{\sigma_{n_D}^2}{\frac{P_r}{K}}  }}\right), 
\end{align} 
where we define $\U\overset{\Delta}=\Uo=\Ut$, $\zeta\overset{\Delta}=\sqrt{\frac{\nu}{\tp P_p}}$, and $\nu\overset{\Delta}=\nu_1=\nu_2$ for notational simplicity. This result implies that exploiting the long-term statistics of the channel, leads to a diversity gain $\mathrm{Tr}\left(\U\right)$ that linearly scales the SINR. On the other hand, only the inverse of uplink SNR $\frac{\sigma_{n_R}^2}{P_u}$, and downlink SNR per user $\frac{\sigma_{n_D}^2}{\frac{P_r}{K}}$ contribute as AWGN to the spectral efficiency. In addition, the impact of channel estimation errors simply appears as $2K\sqrt{\frac{\nu}{\tp P_p}}$ which can be controlled by the pilot specifications and the level of water-filling.  

All in all, we developed a correlation-based analog beamformer followed by simple linear digital processors, i.e., MRC/MRT and ZF scheme. Then, we provided mathematical closed-form expressions which only involve the long-term characteristics of propagation channels. These results reveal that even in the worst-case scenario that multiplexing gain is restricted due to the limited number of RF chains, and array gain is also confined due to the less control of the DSP unit on the analog beamformers, the proposed system can still leverage a substantial diversity gain. This diversity gain is related to the statistical properties of the channels, where analog beamformers exploit the strongest eigenmodes of the channel covariance matrix. This analysis also shows that the dominant statistical terms, i.e., $\mathrm{Tr}\big(\mathbf{U}_i\big)$ for $i=1,2$ have a tendency to boost the desired signal. On the other hand, the moderate statistical terms like $\big\|\mathbf{U}_i\big\|$ for $i=1,2$ mainly contribute to the aggregated noise and interference.   
\subsection{Estimate of Spatial Channel Covariance Matrix}
\label{Channel Covariance}
So far, we have assumed that the relay station knows the true channel covariance matrices; however, this assumption may not hold in practice, and only an estimate of the covariance matrix can be obtained at the relay station. Nevertheless, to properly design an analog beamformer, an accurate spatial channel covariance model is of paramount importance, and should not be overlooked. The estimation of channel covariance matrix is even more controversial in the context of massive MIMO due to the pilot contamination in cellular networks. In this light, there exist several research efforts to address this issue \cite{neumann2017joint,decurninge2015channel}. In \cite{neumann2017joint}, the authors introduce a joint covariance estimation and pilot allocation to reduce the number of pilot symbols during the training phase. In \cite{decurninge2015channel}, the covariance information of the downlink channel is estimated based on the observed covariance of the uplink channel, and without requiring continuous covariance feedback. This novel estimator relies on a dictionary of uplink/downlink pairs of covariance matrices and an interpolation over the Riemannian space. However, these estimators are basically designed for the fully digital massive MIMO systems, and they are not necessarily adoptable to the concept of hybrid configuration paradigm. On the other hand, most existing literature in the field of hybrid massive MIMO systems are tailored to the millimeter wave channels, where the channel sparsity is widely used to estimate the channel covariance matrix \cite{haghighatshoar2017massive,park2016spatial,park2017exploiting,lee2016channel,alkhateeb2014channel}. To the best of our knowledge, the estimation of channel covariance is an interesting topic which is still open for hybrid A/D systems in a rich scattering propagation channel.

{\color{black}
In this paper, we assume that the channel statistics are invariant over the time interval $T_s$ and system bandwidth $B_s$. On the other hand, we note that the instantaneous propagation channel is static over time-frequency blocks of coherence time interval $T_c$ and coherence bandwidth $B_c$. Therefore, $\tc=B_cT_c$ and $\ts=B_sT_s$ denote the numbers of channel uses within which the short-term  and long-term properties of the channel are respectively invariant with $\tc\ll \ts$. As illustrated in Fig. \ref{Block0}, the BS devotes $N_Q$ coherence blocks out of $\lceil{\frac{\ts}{\tc}\rceil}$ possible coherence blocks to estimate the channel covariance matrix. Note that the analog beamformers are not constant over the green areas in these $N_Q$ coherence blocks, though they are invariant at the rest of blocks. Nevertheless, we still consider that they should slowly adopt themselves to the quick variations of the channels due to their analog nature. Hence, it is of uttermost importance to design a covariance estimator to involve few pilot transmissions. Now, let us recall the covariance estimation technique in \cite{pan2017framework,pan2017efficient} which results in the following signal model within the covariance pilot training period (green area):\footnote{{\color{black}We restrict our attention to estimate $\Ro$, however, the same procedure can be similarly pursued for $\Rt$.}}
\begin{align}
\yc=\sqrt{P_p}\Fc\go+\Fd\nc,
\end{align}
where $\go \in \mathbb{C}^{N\times 1}$ denotes the user's channel, $\Fc=\big[\mathbf{F}_1^T[1],\mathbf{F}_1^T[2],\ldots,\mathbf{F}_1^T[\frac{N}{K_a}]\big]^T\in \mathbb{C}^{N \times N}$ is a square matrix collecting the analog beamformer matrices in $\frac{N}{K_a}$ different snapshots. Without loss of generality, we assume that $\frac{N}{K_a}$ is an integer. In addition, $\Fd=\mathrm{blkdiag}\big\{\mathbf{F}_1[1],\mathbf{F}_1[2],\ldots,\mathbf{F}_1[\frac{N}{K_a}]\big\}\in \mathbb{C}^{N \times \frac{N^2}{Ka}}$ denotes a block diagonal matrix, $\nc=\big[\mathbf{n}^T[1],\mathbf{n}^T[2], \ldots,\mathbf{n}^T[{\frac{N}{K_a}}]\big]^T\in \mathbb{C}^{\frac{N^2}{Ka}\times 1}$ stacks the noise vectors, and $\yc=\big[\mathbf{y}^T[1],\mathbf{y}^T[2], \ldots,\mathbf{y}^T[\frac{N}{K_a}]\big]^T\in \mathbb{C}^{N\times 1}$ shows the observation vectors. Also, we note that a discrete Fourier transform (DFT) matrix can be easily chosen to implement the matrix $\Fc$. After some standard manipulations, it is easy to determine the channel covariance matrix $\Ro=\mathbb{E}\big[\go\go^H\big]$ based on the covariance of the baseband received signal vectors $\Rc=\mathbb{E}\big[\yc\yc^H\big]$ as follows \cite{pan2017framework,pan2017efficient}
\begin{figure}[t!]
\centering
		\includegraphics[width=3.40in]{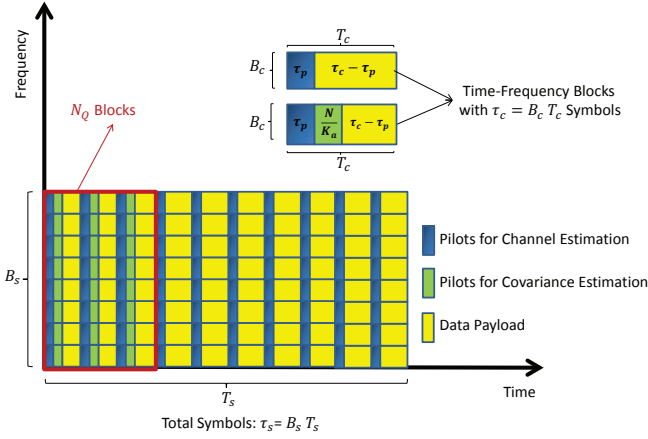}
		\caption{{\color{black}Signal framework with a fixed covariance matrix.}}
		\label{Block0}
\end{figure}
\begin{align}
\label{bridge0}
\Ro=\frac{1}{P_p}\Fc^{-1}\Big(\Rc-\sigma_{n}^2\Fd\Fd^H\Big)\Big(\Fc^H\Big)^{-1},
\end{align}    
\begin{table*}[b!]
\hspace{1cm} \begin{tabular}{ |p{4cm}|p{3cm}||p{3.2cm}|p{4.6cm}|  }
\hline
Parameter& Value &Parameter&Value\\
\hline
Number of antennas & $N=128$            & Coherence time     & $\tc=196$ symbols  						     \\ 
Number of RF chains& $K_a=K_b=50$       & Pilot length       & $\tp=20$ symbols	\\ 
Number of users    & $K=10$             & Mean AoA, AoD      & $\theta_r=\theta_t=0.4\pi$ radian            \\
Relay power        & $P_r=10^{0.5}$  W  & Angle spread       & $\sigma_r=\sigma_t=\sigma=0.25$   radian     \\
User's power       & $P_u=1$         W  & Noise variance		 & $\sigma^2_{n_R}=1$     						          \\
Pilot power        & $P_p=1$         W  & Noise variance		 & $\sigma^2_{n_D}=1$                           \\
\hline
\end{tabular} 
\caption{Simulation parameters.}
\label{table:1} 
\end{table*}
\noindent \!\!\!\! where $\sigma_{n}^2$ denotes the noise power. The equation (\ref{bridge0}) implies that an accurate estimate of $\Rc$ is highly appealing for the better estimation of $\Ro$, i.e. $\hat{\mathbf{R}}_1$. The authors in \cite{pan2017framework,pan2017efficient} suggest the sample covariance matrix $\Rc^{\mathrm{(sample)}}=\frac{1}{N_Q}\sum_{i=1}^{N_Q}\yc\yc^H$ to estimate $\Rc$. However, there are some formidable issues regarding this estimator. First of all, a sample covariance estimator is known as a non-robust technique when the number of samples $N_Q$ is not much larger than the dimension of vector $\yc$ which often occurs in the context of massive MIMO systems. Second, for large dimensional covariance matrices like $\Rc \in \mathbb{C}^{N \times N} $, the sample covariance matrix $\Rc^{\mathrm{(sample)}}$ is typically not well-conditioned and may not even be invertible as its rank is upper limited by the number of samples $N_Q$. This can trigger some bad conditions for other scenarios involving the matrix inversion. Third, the estimation of a covariance matrix with accurate eigenvalues and eigenvectors in massive MIMO is highly challenging. This is due to the fact that the error of all $N^2$ elements can collectively deteriorate the eigenstructures \cite{Bjornson2016massiveMIMO}. Furthermore, the estimation of all $N^2$ entries induces unaffordable levels of pilot training which has deleterious affects on the system spectral efficiency. In this light, we regularize the sample covariance matrix and suggest the following covariance estimator according to \cite{ledoit2004well}
\begin{align}
\Rch=\frac{\beta^2}{\delta^2}\mu\mathbf{I}_N+\frac{\alpha^2}{\delta^2}\Rc^{\mathrm{(sample)}},
\end{align}
where the parameters are obtained in the quartic mean limit, assuming $N$ and $N_Q$ are large enough (about $20$), but their ratio $\frac{N}{N_Q}$ is upper bounded \cite{ledoit2004well}. Then,
\begin{figure}[t!]
\centering
		\includegraphics[width=3.4in]{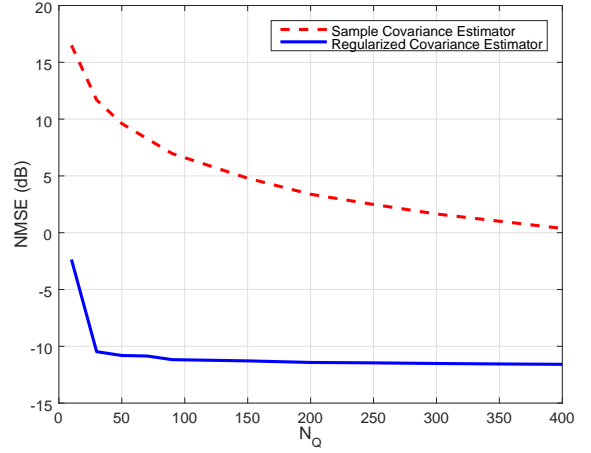}
		\caption{{\color{black}Performance comparison of sample and regularized estimators with $N_Q$ snapshots of covariance matrix $\Ro$ according to the covariance matrix model in (\ref{Covariance_Matrix_Model}) ($N=120$, $K_a=20$, $\sigma_r=\sigma_t=0.6$, $\theta=\frac{\pi}{4}$, SNR=$0$ dB)}.}
		\label{NMSE_R0}
\end{figure}
\begin{align}
\mu&\stackrel{\mathrm{q.m.}}{\longrightarrow}\frac{1}{N}{\mathrm{Tr}\left(\Rc^{\mathrm{(sample)}}\right)},\\
\delta^2&\stackrel{\mathrm{q.m.}}{\longrightarrow}\frac{1}{N}\Big\|\Rc^{\mathrm{(sample)}}-\mu \mathbf{I}_N\Big\|^2,\\
\beta^2&\stackrel{\mathrm{q.m.}}{\longrightarrow}\frac{1}{NN_Q^2}\sum_{i=1}^{N_Q}\Big\|\yc\yc^H-\Rc^{\mathrm{(sample)}}\Big\|^2,\\
\alpha^2&\stackrel{\mathrm{q.m.}}{\longrightarrow}\delta^2-\beta^2.
\end{align}  
\begin{figure*}[b!]
\centering
\begin{subfigure}{.5\textwidth}
  \centering
  \includegraphics[width=3.4in]{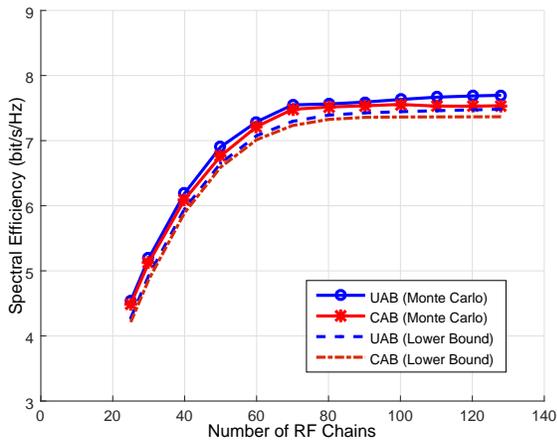}
  \caption{Analog beamformer followed by MRC/MRT DSP unit.}
  \label{UAB_CAB_MRC}
	\end{subfigure}%
\begin{subfigure}{.5\textwidth}
  \centering
  \includegraphics[width=3.4in]{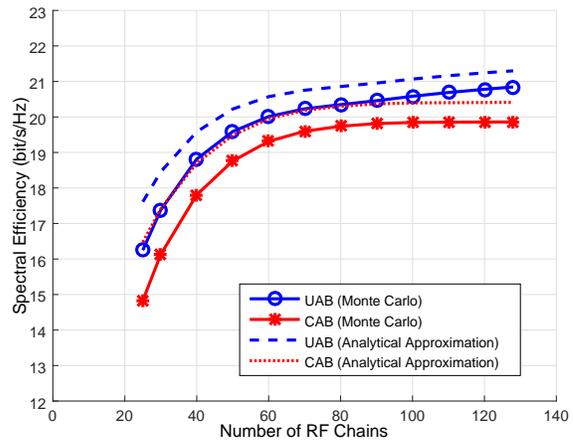}
  \caption{Analog beamformer followed by ZF DSP unit.}
  \label{UAB_CAB_ZF}
\end{subfigure}
	\centering
\caption{Performance assessment of the proposed analog beamformer in the hybrid A/D structure.}
\label{UAB_CAB}
\end{figure*}
\begin{figure*}[b!]
\centering
\begin{subfigure}{0.5\textwidth}
\centering
\includegraphics[width=3.4in]{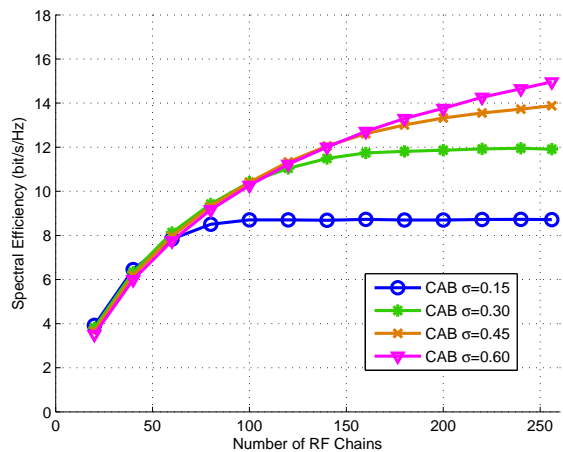}
\caption{MRC/MRT Scheme $N=256$}
\label{MRC_sigma}
\end{subfigure}%
\begin{subfigure}{0.5\textwidth}
\centering
\includegraphics[width=3.4in]{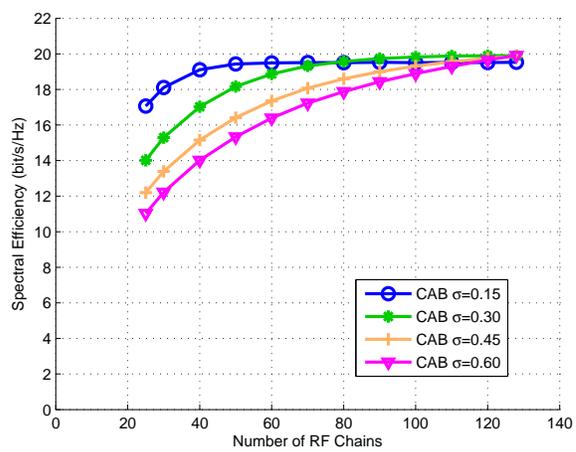}
\caption{ZF Scheme $N=256$}
\label{ZF_sigma}
\end{subfigure}
\caption{Performance of the proposed analog beamformer with different levels of correlation.}
\label{sigma}
\end{figure*}
\begin{figure}[t!]
\centering
		\includegraphics[width=3.4in]{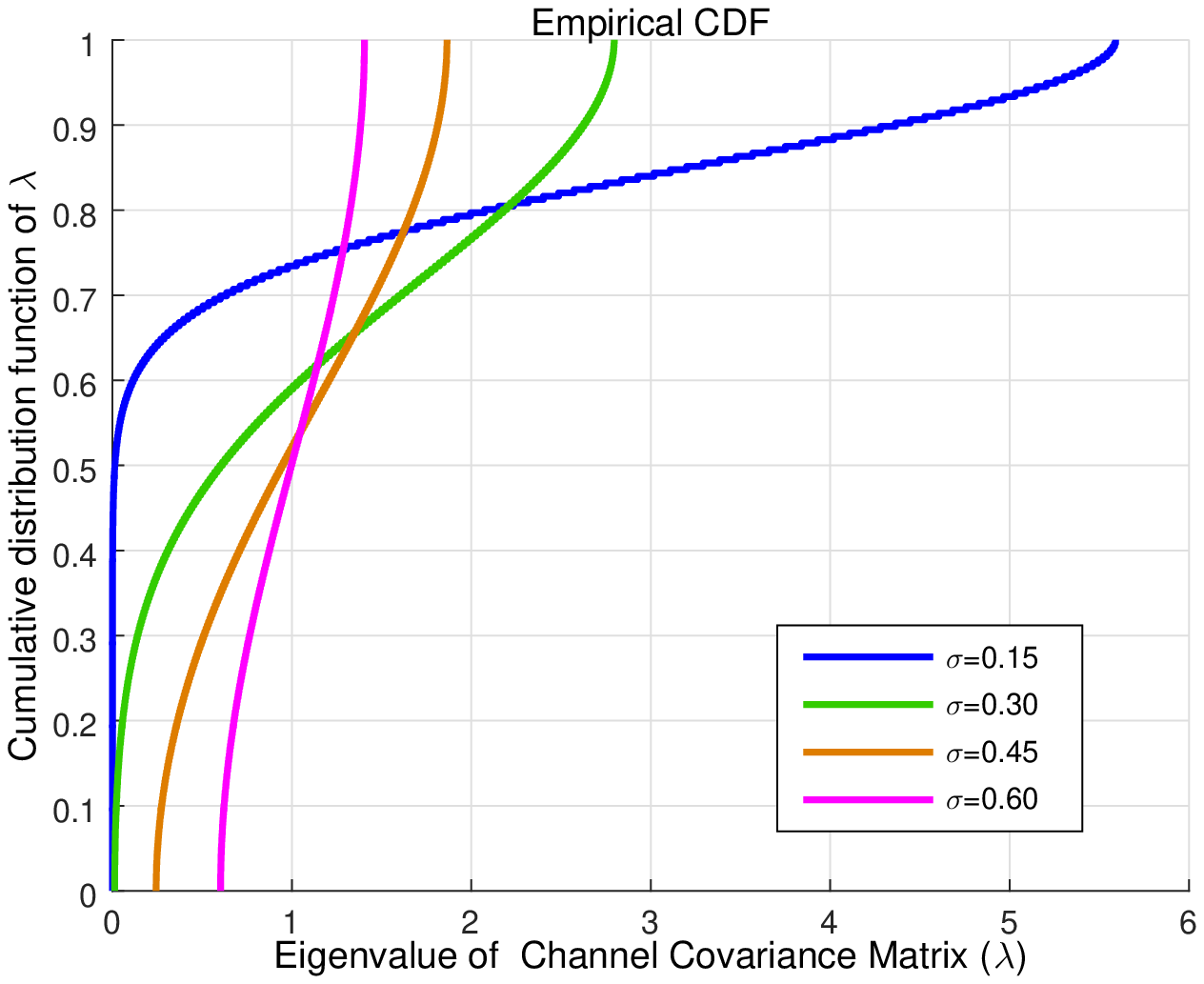}
		\caption{Empirical cumulative distribution function.}
		\label{cdf}
\end{figure}
As we discussed earlier in the manuscript, the statistics of the propagation channel are invariant over a very long time horizon. For instance, a channel with a coherence bandwidth of $B_c=200$KHz and a coherence time of $T_c=1$ms, which allows for $135$Km/h mobility at a carrier frequency of $2$GHz, can typically preserve its long-term statistics for $B_s=10$MHz and $T_s=0.5$s \cite{Bjornson2016massiveMIMO}. Thus, the covariance matrix of this channel can be assumed fixed over $N_R=\frac{\ts}{\tc}=\frac{B_sT_s}{B_cT_c}=25,000$ time-frequency coherence blocks. However, Fig. \ref{NMSE_R0} shows that only $N_Q\approx 25$ is required to achieve an estimate of the covariance matrix with $-11$dB normalized mean square error (NMSE), $\mathbb{E}\left[\frac{\|\Ro-\hat{\mathbf{R}}_1\|^2}{\|\Ro\|^2}\right]$. Note that, with $N_Q=25$, the estimation overhead is only a small portion $\frac{N_Q}{N_R}=0.001$ of the available time-frequency coherence blocks. In this light, one may devote the entire or the major part of each of these $N_Q$ blocks just to estimate the covariance matrix, but with a very slow rate to reduce the variations of analog beamformers. For the design of fast analog beamformers, we refer the interested readers to multiple training technique in \cite{pan2017framework,pan2017efficient}.}

\section{Numerical Results}
\label{Simulation Results Journal}
In this section, we provide numerical results to evaluate the performance of hybrid A/D multipair massive relaying with the channel estimation at the relay station. We model the channel covariance matrices, $\Ro$ and $\Rt$, as follows:
\begin{align}
\label{Covariance_Matrix_Model}
\hspace{-7pt} \left[\Ro\right]_{m,n}\hspace{-4pt}=&\hspace{-1pt}e^{-j2\pi(n-m)\Delta_r\cos(\theta_r)} e^{-\frac{1}{2}\left(2\pi(n-m)\Delta_r\sin(\theta_r)\sigma_r\right)^2}\hspace{-2pt},  \\
\hspace{-7pt} \left[\Rt\right]_{m,n}\hspace{-4pt}=&\hspace{-1pt}e^{-j2\pi(m-n)\Delta_t\cos(\theta_t)} e^{-\frac{1}{2}\left(2\pi(m-n)\Delta_t\sin(\theta_t)\sigma_t\right)^2}\hspace{-2pt},  
\end{align} 
where $\Delta_r$,  $\Delta_t$ denote the antenna spacing, $\theta_r$, $\theta_t$, $\sigma_r^2$, and $\sigma_r^2$ represent the mean angle of arrival (AoA) to the relay station, mean angle of departure (AoD) from the relay station, receive angle spread and transmit angle spread, respectively \cite{ertel1998overview,bolcskei2003impact}. These correlation models basically represent Gaussian matrices with a spread inversely proportional to the product of the antenna spacing and angle spread. We note that a smaller angle spread represents higher level of spatial correlation. It is also noteworthy that this model can be easily expanded to the clustered channels where each cluster corresponds to a specific AoA and AoD, and there are sufficient scatterers in each cluster: approximately $10$ or more \cite{zetterberg1995spectrum}. For the sake of simplicity, we assume that the channel matrices are normalized so that $\sigma^2_{n_R}$ and $\sigma^2_{n_D}$ contain both the noise variance and pathloss. Unless otherwise specified, we utilize the detailed parameter settings as summarized in Table \ref{table:1}.
All analytical results obtained in this paper assume that phase shifters are designed in the digital domain, and they are able to take any modulus and phase. However, as we described earlier in (\ref{Analog Optimization}), this so-called \textit{unconstrained analog beamforming} (UAB) design, is not quite practical due to the fact that analog beamformers are implemented by means of phase shifters with a constant modulus constraint. To circumvent this problem, we first design the phase shifters as before, and then will normalize their modulus to satisfy the aforementioned constraint. For the sake of presentation clarity, we call this methodology as \textit{constrained analog beamforming} (CAB) satisfying (\ref{Analog Optimization}). 

\begin{figure*}[b!]
\centering
\begin{subfigure}{0.5\textwidth}
\centering
\includegraphics[width=3.4in]{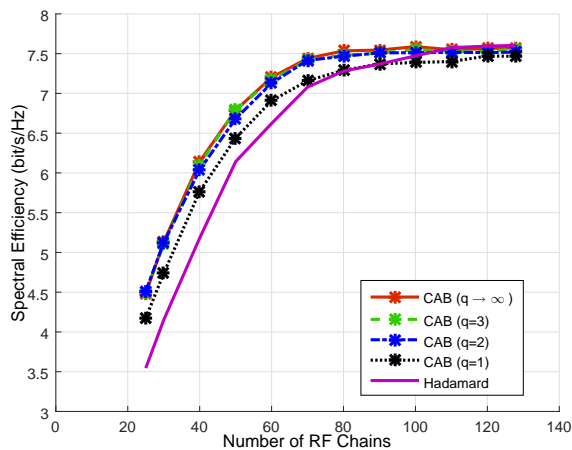}
\caption{MRC/MRT Scheme}
\label{MRC_CAB_Quantization}
\end{subfigure}%
\begin{subfigure}{0.5\textwidth}
\centering
\includegraphics[width=3.4in]{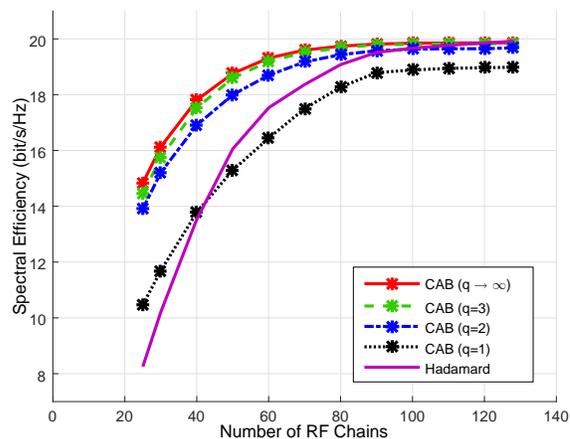}
\caption{ZF Scheme}
\label{ZF_CAB_Quantization}
\end{subfigure}
\caption{Performance of the hybrid A/D configuration with quantized phase shifters.}
\label{CAB_Quantization}
\end{figure*}

Figure \ref{UAB_CAB} shows a great congruency between Monte-Carlo simulation and our analytical result. Moreover, it showcases that the gap between unconstrained (UAB) and constrained analog beamforming (CAB) design is negligible. This observation implies that our methodology is very robust with respect to removing the modulus constraint of the phase shifters. In addition, ZF processing can greatly enhance the spectral efficiency compared to the MRC/MRT processing by nulling out the inter-user interference. This signifies that by deploying only $50$ RF chains the hybrid A/D configuration paradigm can capture more than $90 \%$ of the spectral efficiency offered by the fully digital structure with $128$ RF chains.  

Fig. \ref{sigma} illustrates the performance of hybrid A/D topology with different levels of channel spatial correlation. Fig. \ref{ZF_sigma} infers that with a limited number of RF chains, the proposed analog beamformer followed by ZF scheme can harness channel correlation. It is due to this fact that, in highly correlated channels, i.e., smaller $\sigma$, the eigenvalues of channel covariance matrix are decentralized and will be spread on a wider range (See Fig. \ref{cdf}). Therefore, there exist a few strong modes (paths) that can be captured by the limited available number of RF chains. On the other hand, MRC/MRT scheme is unable to decorrelate and extract the strong eigenmodes. Thus, Fig. \ref{MRC_sigma} showcases a poor performance with limited number of RF chains. 

\begin{figure}[t!]
\centering
   \includegraphics[width=3.4in]{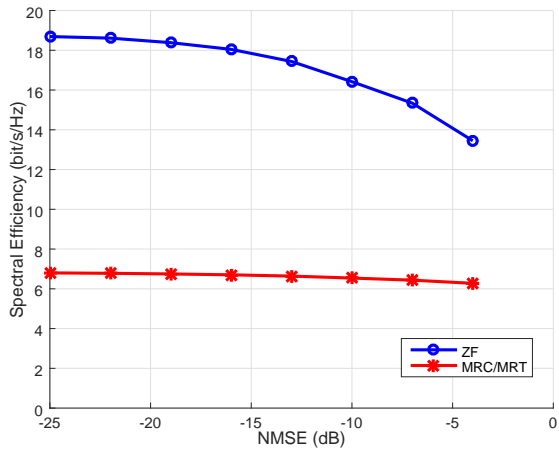}
  \caption{Performance of the hybrid A/D structure for different NMSE of covariance matrix.}
	\vspace{-0.5cm}
	\label{NMSE}
\end{figure}

Figure \ref{CAB_Quantization} reveals that the hybrid A/D configuration paradigm is robust to phase shifters with coarse quantization. This is due to the fact that only part of the signal processing is performed in the analog domain and the rest of signal processing is passed to the DSP unit in the baseband domain. To be accurate, numerical results showcase that the analog beamformer is robust to phase quantization, and even with $2$-bit resolution, the system can capture more than $93 \%$ of the spectral efficiency offered by the ideal phase shifters. In addition, Fig. \ref{CAB_Quantization} compares the performance of our analog beamformer with $1$-bit resolution with current existing binary-valued benchmarks, i.e., discrete Hadamard transform beamformer \cite{kim2015mse}. It demonstrates that our proposed analog beamformer outperforms the Hadamard beamformer, particularly, for limited number of RF chains which suits to the hybrid A/D structure. It is noteworthy that the hybrid A/D structure offers a high spectral efficiency with ZF, but is less sensitive to coarse quantization with the MRC/MRT scheme.
\begin{figure*}[t!]
\centering
\begin{subfigure}{.5\textwidth}
  \centering
  \includegraphics[width=3.4 in]{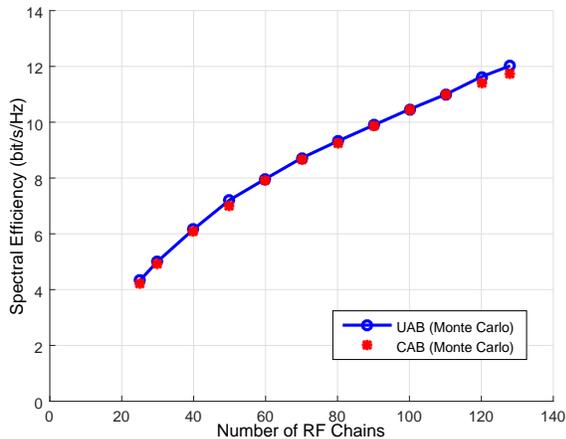}
  \caption{Analog beamformer followed by MRC/MRT DSP unit.}
  \label{AntennasRatio_MRC}
	\end{subfigure}%
\begin{subfigure}{.5\textwidth}
  \centering
  \includegraphics[width=3.4 in]{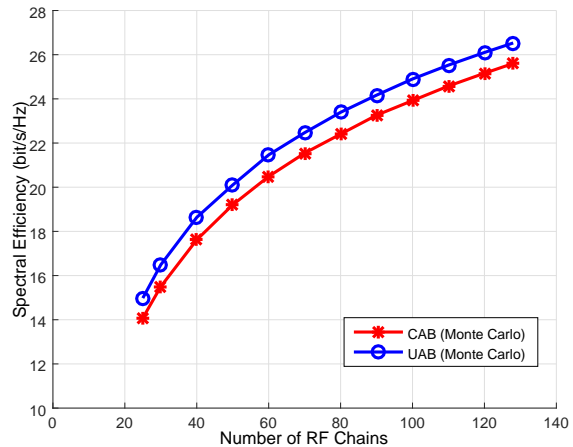}
  \caption{Analog beamformer followed by ZF DSP unit.}
  \label{AntennasRatio_ZF}
\end{subfigure}
	\centering
\caption{The impact of the number of relay antennas with $\frac{N}{K_a} = 3$.}
\label{AntennasRatio}
\end{figure*}
\begin{figure*}[t!]
\centering
\begin{subfigure}{.5\textwidth}
  \centering
  \includegraphics[width=3.4in]{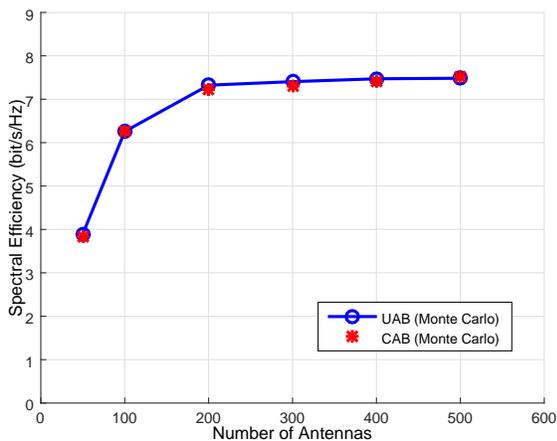}
  \caption{Analog beamformer followed by MRC/MRT DSP.}
  \label{Diversity_Gain_MRC}
\end{subfigure}%
\begin{subfigure}{.5\textwidth}
  \centering
  \includegraphics[width=3.4in]{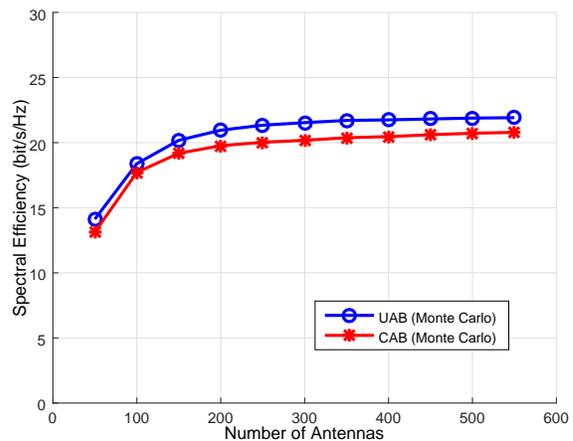}
  \caption{Analog beamformer followed by ZF DSP.}
  \label{Diversity_Gain_ZF}
	\end{subfigure}
	\centering
\caption{The impact of the number of relay antennas with constant number of RF chains $K_a=K_b=50$.}
\label{Antennas}
\end{figure*}
Figure \ref{AntennasRatio} evaluates the impact of deployed antennas at the relay station, where we increase the number of antennas while the ratio $N/K_a$ is kept constant.  The more antennas we deploy at the relay station, the more diversity gain we can achieve. Therefore, the spectral efficiency will increase, although the ratio of the service antennas and RF chains is still constant. More interestingly, Fig. \ref{Antennas} demonstrates that the system performance improves by deploying more antennas, even though the number of RF chains is fixed $K_a=K_b=50$. This is due to the fact that by deploying more antennas, the number of eigenvalues of the channel covariance matrices ($\Ro$, $\Rt$) will increase, and the proposed analog beamformers smartly collect only the strongest ones. In other words, the water-filling algorithm enhances the system performance by exploiting the diversity gain from the long-term  characteristics of propagation channels.   
Figure \ref{NMSE} illustrates the robust performance of the proposed hybrid structure against the NMSE of covariance matrix estimation. Although MRC/MRT scheme is more robust to the covariance matrix estimation, its spectral efficiency is still more than $2$ times less than the ZF scheme, even for higher NMSE.  

\begin{figure}[t!]
\centering
   \includegraphics[width=3.4in]{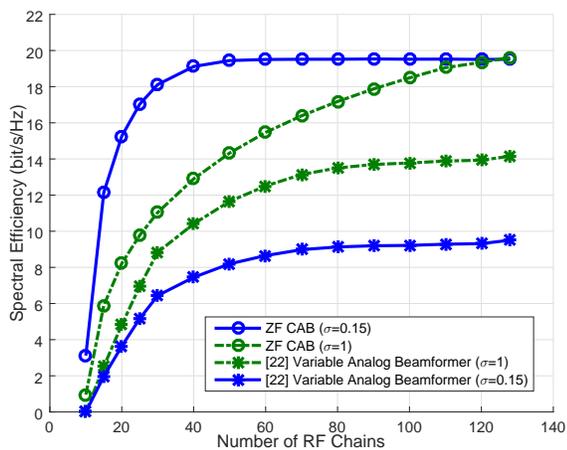}
  \caption{Performance of the proposed hybrid A/D structure compared to the hybrid strategy of \cite{xu2017spectral} equipped with variable analog beamformers.}
	\vspace{-0.5cm}
	\label{ZF_ShiJin}
\end{figure}

\begin{figure}[t!]
\centering
   \includegraphics[width=3.4in]{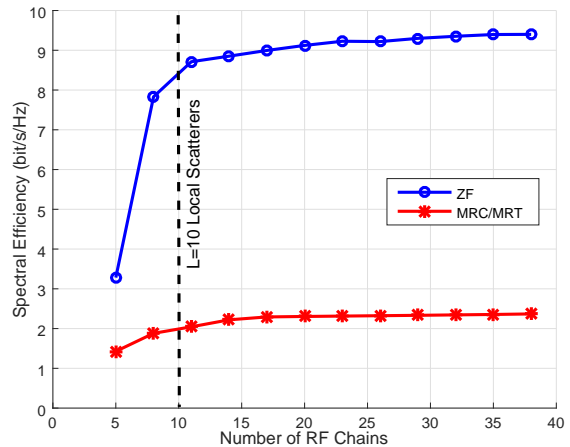}
  \caption{Performance of the hybrid A/D structure under a parametric channel model $\big(N=64$, $K=5$, $\tp=10$, $d=\frac{\lambda}{2}$, $L=10$, and $\phi_l\sim U\left[0,2\pi\right)\big)$.}
	\label{Parametric_Channel}
\end{figure}

Performance of the proposed method with the hybrid strategy in \cite{xu2017spectral} is presented in Fig. \ref{ZF_ShiJin}. In \cite{xu2017spectral}, the analog beamformers are designed in each channel coherence block, and therefore, phase shifters should be adapted to the quick variations of the propagation channels which is, of course, a challenging task due to the low flexibility of analog beamformers. In addition, the acquisition of full CSI at the relay station comes at the price of severe signaling overhead which scales badly with the number of antennas, i.e., $2K\lceil\frac{N}{\min\left(K_a,K_b\right)}\rceil$ pilot symbols per channel coherence block. This extra overhead signaling in each channel coherence block imposes a critical limitation, particularly, for high mobility scenarios and higher frequencies. On the other hand, our proposed beamformer depends on the long-term statistics of the channel which varies $100-1000$ times slower than the small-scale fading. Furthermore, Fig. \ref{ZF_ShiJin} reveals that our correlation-based hybrid strategy outperforms \cite{xu2017spectral} in terms of spectral efficiency and this result is more pronounced for highly correlated channels, e.g., lower $\sigma$.
Finally, borrowing the channel model from \cite{xu2017spectral}, we investigate the performance of our hybrid A/D beamformer under the parametric channel model
\begin{align}
\gok=\sqrt{\frac{N}{L}}\sum_{l=1}^{L}\alpha_k^l \mathbf{a}_{r}\left(\phi_l\right) \label{g_k},
\end{align}
where $L$ denotes the number of scatterers, and consequently, number of paths toward the relay station. The random variable $\alpha_k^l\sim\mathcal{CN}\left(0,1\right)$ represents the complex gain of the $l$-th path, and $\mathbf{a}_{r}\left(\phi_l\right)$ is the antenna array response which can be formulated for the uniform linear arrays as 
\begin{align}
\mathbf{a}_{r}\hspace{-2pt}\left(\phi_l\right)\hspace{-3pt}=\hspace{-3pt}\frac{1}{\sqrt{N}}\Big[1,e^{j\frac{2\pi d}{\lambda}\hspace{-2pt}\cos\left(\phi_l\right)}\hspace{-2pt}, \ldots,e^{j\frac{2\pi(N-1)d}{\lambda}\hspace{-2pt}\cos\left(\phi_l\right)}\hspace{-2pt}\Big],
\end{align}
where $\lambda$ denotes the signal wavelength, $d$ represents the distance between two adjacent relay antennas, and $\phi_l$ is the angle of incidence of the line-of-sight for the $l$-th path onto the receive antenna array. Expressing (\ref{g_k}) in a simpler form, one has  
$
\gok=\sqrt{\frac{N}{L}}\mathbf{A}_r\hok,
$
where $\mathbf{A}_r=\big[\mathbf{a}_{r}\left(\phi_1\right),\mathbf{a}_{r}\left(\phi_2\right), \ldots, \mathbf{a}_{r}\left(\phi_L\right)\big]$, and $\hok=\left[\alpha_k^1,\alpha_k^2,\ldots,\alpha_k^L\right]^T$. Now, taking all the users into account we can model the propagation channel $\Go$ as $ \Go=\sqrt{\frac{N}{L}}\mathbf{A}_r\Ho$, 
where $\Ho=\big[\mathbf{h}_{1,1},\mathbf{h}_{1,2},\ldots,\mathbf{h}_{1,K}\big]$. Here, we note that the rank of $\Go$ is restricted to the number of local scatterers $L$ close to the relay station. The same propagation model can be similarly deduced for the other channel $\Gt$. In Fig. \ref{Parametric_Channel}, we showcase the spectral efficiency of the proposed hybrid A/D beamforming under the parametric channel model with azimuth angle $\phi_l$ chosen from a uniform distribution over $\left[0, 2\pi\right)$ which reveals that only $10$ RF chains, in accordance to $L=10$ propagation paths around the relay station, is enough to achieve most of the spectral efficiency.
\section{Conclusion}
\label{Conclusion Journal}
Massive MIMO is a promising technique for the next generation of wireless communication systems which addresses most of the critical challenges associated with concurrent relaying systems, such as DSP complexity, long processing delay, and low SINR at the cell edges. However, massive relaying experiences high fabrication/implementation cost and power consumption due to the large number of RF chains. To overcome this limitation, we proposed to reduce the number of RF chains in a viable analog-digital configuration which is usually referred to hybrid A/D structure. It is well-known that this structure reduces the multiplexing gain and also restricts the power gain due to the reduced flexibility of analog beamformers, particularly, with imperfect CSI at the relay station. Assuming a correlated fading channel, we designed a novel analog beamformer which exploits the long-term channel statistics and results in high spectral efficiency and robustness with respect to phase quantization. We also derived an approximation and a lower bound on the spectral efficiency which involves the long-term parameters of the propagation channels. 
\appendices
\section{Prerequisite Lemmas}
\label{Prerequisite Lemmas}
\begin{Lemma1_MRC}
\label{Lemma1_MRC}
Let $\mathbf{H} \in \mathbb{C}^{N \times K}$ be a zero-mean Gaussian random matrix with i.i.d. entries with variance $\sigma^2$. Also, assume that \mbox{$\mathbf{V} \in \mathbb{C}^{N \times N}$} is a deterministic matrix. Then, we have
\begin{align}
\mathbb{E}\big[\mathbf{H}^H\mathbf{V}\mathbf{H}\big]=\sigma^2\mathrm{Tr}\left(\mathbf{V}\right)\mathbf{I}_K.
\end{align}
\end{Lemma1_MRC}
\begin{Lemma2_MRC}
\label{Lemma2_MRC}
Let the vectors $\mathbf{h}$, $\mathbf{g} \in \mathbb{C}^N$ be two independent zero-mean circular Gaussian random vectors such that $\mathbf{h}$, $\mathbf{g} \sim \mathcal{CN}\left(0, \mathbf{I}_N\right)$, and consider the deterministic matrix $\U$, then it holds that 
\begin{align}
\mathbb{E}\bigg[\Big|\mathbf{h}^H\U\mathbf{g} \Big|^2\bigg]=&\big\|\U\big\|^2,\\
\mathbb{E}\bigg[\Big|\mathbf{h}^H\U\mathbf{h} \Big|^2\bigg]=&\Big|\mathrm{Tr}\big(\U\big)\Big|^2 + \big\|\U\big\|^2. \label{Emil Formula}
\end{align}
\begin{proof}
See \cite[Lemma 2]{bjornson2015massive}.
\end{proof}
\end{Lemma2_MRC}
\begin{Lemma3_MRC}
\label{Lemma3_MRC}
Let $\G=\U^\frac{1}{2}\mathbf{H}$, where $\mathbf{H}\in \mathbb{C}^{N \times K}$ is a zero-mean Gaussian random matrix with i.i.d. entries of unit variance. Then, for any Hermitian deterministic matrix $\mathbf{U} \in \mathbb{C}^{N \times N}$ we have
\begin{align}
\mathbb{E}\Big[\G^H \G \G^H \G\Big]=\Big(\mathrm{Tr}^2\big(\U\big) + K\big\|\U\big\|^2\Big)\mathbf{I}_K.
\end{align}
\end{Lemma3_MRC}
\begin{proof}
Let us define $\Q \stackrel{\Delta}{=}\mathbb{E}\Big[{\G^H \G \G^H \G}\Big]$. Then, the $\left(i,i\right)$-th element of matrix $\Q$ is given by
\begin{align}
\big[\Q\big]_{i,i}=&\mathbb{E}\Big[\gi^H\Big(\displaystyle\sum_{m=1}^{K}{\gm\gm^H}\Big)\gi\Big] \nonumber \\
=&\mathbb{E}\Big[\gi^H\gi\gi^H\gi+\gi^H\Big(\displaystyle\sum_{\substack{m \neq i}}^{K}{\gm\gm^H}\Big)\gi\Big] \nonumber \\
=&\mathbb{E}\Big[\big|\gi^H\gi\big|^2\Big] + \displaystyle\sum_{\substack{m \neq i}}^{K}\mathbb{E}\Big[\big|\gi^H\gm \big|^2\Big] \nonumber \\
=&\mathrm{Tr}^2\left(\U\right) + K\left\|\U\right\|^2.
\end{align}
We note that in the last equation we applied Lemma \ref{Lemma2_MRC}. In a similar spirit we can show the off-diagonal elements of matrix $\Q$ are zero. This concludes the proof.
\end{proof}
\begin{Lemma4_MRC}
\label{Lemma4_MRC}
Let the vector $\mathbf{h} \in \mathbb{C}^N$ be a zero-mean circular Gaussian random vector such that $\mathbf{h}\sim\mathcal{CN}\left(0,\mathbf{I}_N\right)$. Also let us assume that $\U \in \mathbb{C}^{N \times N}$ is a positive definite matrix with a limited spectral norm, $\lambda_{\mathrm{max}}\big( \U \big) < \infty$, where $\lambda_{\mathrm{max}}\big( \U \big)$ denotes the largest eigenvalue of matrix $\U$. Then, it can be shown that for $N \to \infty$  
\begin{align}
\frac{\mathbf{h}^H\U\mathbf{h}}{\mathrm{Tr}\big(\U\big)}\stackrel{\mathrm{a.s.}}{\longrightarrow}1.
\end{align}
\end{Lemma4_MRC}
\begin{proof}
This Lemma can be directly proved by invoking the Chebyshev's inequality for the random variable $X=\frac{\mathbf{h}^H\U\mathbf{h}}{\mathrm{Tr}\big(\U\big)}$ 
\begin{align}
\mathrm{Pr}\bigg(\Big|X-\mathbb{E}\left[X\right]\Big|\geq \epsilon \bigg) \leq \frac{\mathrm{var}\left(X\right)}{\epsilon^2},
\end{align}
where we assume that $\epsilon$ is a very small real number. The expectation and power of the random variable $X$ are given by (see equation (\ref{Emil Formula})) 
\begin{align}
\mathbb{E}\Big[X\Big]&=1,\\
\mathbb{E}\Big[X^2\Big]&=\frac{\mathrm{Tr}^2\big(\mathbf{U}\big)+\big\|\mathbf{U}\big\|^2}{\mathrm{Tr}^2\big(\mathbf{U}\big)}.
\end{align}
Let us assume that $\lambda_i$ denotes the $i$-th biggest eigenvalue of matrix $\U$, then
\vspace{-0.5cm}
\begin{align}
\mathrm{var}\left(X\right)=&\frac{\big\|\mathbf{U}\big\|^2}{\mathrm{Tr}^2\big(\mathbf{U}\big)} = \frac{\displaystyle\sum_{i=1}^{N}{\lambda_i^2}}{\displaystyle\sum_{i=1}^N{\lambda_i^2}+\displaystyle\sum_{\substack{i=1 \\ i \neq j}}^N \displaystyle\sum_{j=1}^N\lambda_i\lambda_j} \nonumber \\
\leq & \frac{\displaystyle\sum_{i=1}^{N}{\lambda_i^2}}{\displaystyle\sum_{i=1}^N{\lambda_i^2}+(N^2-N)\lambda_N}\leq \frac{1}{1+\frac{\left(N-1\right)\lambda_N^2}{\lambda_1^2}},
\label{Cheby}
\end{align}
and considering that $\lambda_1< \infty$ we can conclude that for any positive real number $\epsilon$ 
\begin{align}
\lim_{N \to \infty} \frac{\mathrm{var}\left(X\right)}{\epsilon^2}=0,
\end{align} 
which implies that
\begin{align}
\lim_{N \to \infty}\mathrm{Pr}\bigg( \Big|X-\mathbb{E}\left[X\right]\Big|\geq \epsilon \bigg) = 0.
\end{align}
Thus, the proof is complete.
\end{proof}
\begin{Lemma5_MRC}
\label{Lemma5_MRC}
Let the vectors $\mathbf{h}$, $\mathbf{g} \in \mathbb{C}^N$ be two independent zero-mean circular Gaussian random vectors such that $\mathbf{h}$, $\mathbf{g} \sim \mathcal{CN}\left(0, \mathbf{I}_N\right)$. Also, let us assume that $\U \in \mathbb{C}^{N \times N}$ is a positive definite matrix with a limited spectral norm. Then, it can be shown that for $N \to \infty$  
\begin{align}
\frac{\mathbf{h}^H\U\mathbf{g}}{\mathrm{Tr}\big(\U\big)}\stackrel{\mathrm{a.s.}}{\longrightarrow}0.
\end{align}
\end{Lemma5_MRC}
\begin{proof}
The proof is in a similar manner as Lemma {\ref{Lemma4_MRC}}, by defining a zero-mean random variable $Y=\frac{\mathbf{h}^H\U\mathbf{g}}{\mathrm{Tr}\big(\U\big)}$ and considering that
\vspace{-0.5cm}
\begin{align}
\mathrm{var}\big(Y\big)&=\frac{\big\|\mathbf{U}\big\|^2}{\mathrm{Tr}^2\big(\mathbf{U}\big)}. 
\end{align}  
\end{proof}
\vspace{-0.4cm}
\begin{Corollary1_MRC}
\label{Corollary1_MRC}
Let $\mathbf{H} \in \mathbb{C}^{K \times N} $ be a Gaussian random matrix with i.i.d $\mathcal{CN}\left(0,\mathbf{I}_N\right)$ entries. Also let us assume that $\U \in \mathbb{C}^{N \times N}$ is a positive definite matrix with a limited spectral norm. Then, it can be shown that for $N \to \infty$  
\begin{align}
\frac{\mathbf{H}^H\U\mathbf{H}}{\mathrm{Tr}\big(\U\big)}\stackrel{\mathrm{a.s.}}{\longrightarrow} \mathbf{I}_K.
\end{align}
\end{Corollary1_MRC}
\noindent \textit{Remark $2$}: Lemmas \ref{Lemma4_MRC}, \ref{Lemma5_MRC} and consequently Corollary \ref{Corollary1_MRC} remain agnostic of semi-definite (and ill-conditioned) matrices with large number of non-zero (and non-trivial) eigenvalues. To prove this, just define the number of non-trivial and non-zero eigenvalues of matrix $\U$ as $N ^\prime$. Then, the proof follows trivially by substituting $N ^\prime$ instead of $N$ in (\ref{Cheby}). 
\vspace{-0.5cm}
\section{Proof of Proposition \ref{Term1_MRC}}
\label{Term1_MRC_ap}
The proof of the first part is trivial by applying Lemma \ref{Lemma1_MRC} in Appendix \ref{Prerequisite Lemmas}. For the second part let us define an auxiliary variable namely $t^{\mathrm{mrc}}_{\mathrm{aux}}$, and then simplify it by expanding the matrix product around its columns. Then, considering this fact that the columns of the estimation channels are independent, and by recalling Lemma \ref{Lemma2_MRC} we have 

\begin{align}
t^{\mathrm{mrc}}_{\mathrm{aux}}&=\am^2\mathbb{E}\bigg[\Big|\gthk^H\Gth\Goh^H\gohk \Big|^2\bigg] \nonumber \\
&=\am^2\mathbb{E}\Bigg[\bigg|\gthk^H \Big(\sum_{m=1}^K{\gthm\gohm^H}\Big)\gohk\bigg|^2\Bigg] \nonumber \\ 
&=\am^2\sum_{m=1}^K{\mathbb{E}\Bigg[\bigg|\gthk^H \gthm\gohm^H\gohk\bigg|^2\Bigg]}\nonumber \\
&\overset{\left(\mathrm{L}\ref{Lemma2_MRC}\right)}= \am^2\Big(\mathrm{Tr}^2\left(\Ut\right) + \left\|\Ut\right\|^2\Big) \Big(\mathrm{Tr}^2\left(\Uo\right) + \left\|\Uo\right\|^2\Big) \nonumber \\
& + \am^2(K-1)\big\|\Uo\big\|^2\big\|\Ut\big\|^2.  
\end{align}
The proof is completed by using $t_1^{\mathrm{mrc}}=t^{\mathrm{mrc}}_{\mathrm{aux}}-t^{\mathrm{mrc}}_0$. For the sake of notational simplicity, hereafter, we use $\left(\mathrm{L}\ref{Lemma1_MRC}\right)$, $\left(\mathrm{L}\ref{Lemma2_MRC}\right)$, and $\left(\mathrm{L}\ref{Lemma3_MRC}\right)$ to denote Lemmas $\ref{Lemma1_MRC}$--$\ref{Lemma3_MRC}$.
\vspace{-0.1cm}
\section{Proof of Proposition \ref{Term3_MRC}}
\label{Term3_MRC_ap}
By leveraging the property of Gaussian random matrices from Appendix \ref{Prerequisite Lemmas}, we expand the proof as follows
\begin{align}
t_3^{\mathrm{mrc}}&=\am^2\mathbb{E}\bigg[\Big|\gthk^H\Gth\Goh^H\Eo\mathbf{x}\Big|^2\bigg] \nonumber\\
&= \am^2\mathbb{E}\Big[\gthk^H\Gth\Goh^H\Eo\Eo^H\Goh\Gth^H\gthk\Big] \nonumber \\
&\overset{\left(\mathrm{L}\ref{Lemma1_MRC}\right)}{=} K\am^2\mathbb{E}\Big[\gthk^H\Gth\Goh^H\Ueo\Goh\Gth^H\gthk \Big] \nonumber \\
&\overset{\left(\mathrm{L}\ref{Lemma1_MRC}\right)}{=} K \am^2 \mathrm{Tr}\big(\Uo\Ueo\big)\mathbb{E}\Big[\gthk^H\Gth\Gth^H\gthk \Big] \nonumber\\
&\overset{\left(\mathrm{L}\ref{Lemma3_MRC}\right)}{=} K \am^2 \mathrm{Tr}\big(\Uo\Ueo\big) \Big(\mathrm{Tr}^2\left(\Ut\right) + K\left\|\Ut\right\|^2\Big).  
\end{align}
\vspace{-0.5cm}
\section{Proof of Proposition \ref{Term4_MRC}}
\label{Term4_MRC_ap}
By invoking Lemma \ref{Lemma1_MRC} and Lemma \ref{Lemma3_MRC} we get
\begin{align}
t_4^{\mathrm{mrc}}&= \am^2 \mathbb{E}\bigg[\Big|\etk^H\Gth\Goh^H\Goh \mathbf{x}\Big|^2\bigg] \nonumber \\
&=\am^2 \mathbb{E}\bigg[\etk^H \Gth\Goh^H \Goh \Goh^H \Goh \Gth^H \etk \bigg] \nonumber\\
& \overset{\left(\mathrm{L}\ref{Lemma1_MRC}\right)}{=} \am^2\mathbb{E}\bigg[\mathrm{Tr}\Big(\Uet \Gth\Goh^H \Goh \Goh^H \Goh \Gth^H \Big)\bigg] \nonumber \\
&\overset{\left(\mathrm{L}\ref{Lemma3_MRC}\right)}{=}\am^2\Big(\mathrm{Tr}^2\left(\Uo\right) + K\left\|\Uo\right\|^2\Big)\mathbb{E}\bigg[\mathrm{Tr}\big( \Uet\Gth\Gth^H \big)\bigg] \nonumber \\
& \overset{\left(\mathrm{L}\ref{Lemma1_MRC}\right)}{=} \am^2K \mathrm{Tr}\left( \Ut\Uet \right) \Big(\mathrm{Tr}^2\left(\Uo\right) + K\left\|\Uo\right\|^2\Big). 
\end{align}
\vspace{-0.2cm}
\section{Proof of Proposition \ref{Term1_ZF}}
\label{Term1_ZF_ap}
The proof is straightforward by invoking Corollary \ref{Corollary1_MRC} ($C1$) 
\begin{align}
&\az^2\mathbb{E}\bigg[\Gohi\Goh^H\Eo\mathbf{x} \mathbf{x}^H \Eo^H \Goh \Gohi\bigg] \nonumber \\
&= K\az^2\mathbb{E}\bigg[\Gohi\Goh^H\Ueo\Goh \Gohi\bigg] \nonumber
\end{align}
\begin{align}
\overset{(C\ref{Corollary1_MRC})}{\stackrel{\mathrm{a.s.}}{\longrightarrow}}\frac{K\az^2\mathrm{Tr}\big(\Uo\Ueo\big)}{\mathrm{Tr}^2\big(\Uo\big)}\mathbf{I}_K. 
\end{align}
\vspace{-0.25cm}
\section{Proof of Proposition \ref{Trace Connection}}
\label{Projection}
Without loss of generality, we just prove the proposition for the case $i=1$, and the same result can be similarly deduced for the case $i=2$. Let $\Ro=\Uro\Sro\Uro^H$ represent the eigenvalue decomposition of the channel covariance matrix, then we have
\begin{align}
\label{LAMBDA}
   \Sro=
  \left[ {\begin{array}{cc}
   \Sro^{'} & \mathbf{0} \\
    \mathbf{0} & \Sro^{''} \\
  \end{array} } \right],
\end{align}
where $\Sro=\mathrm{diag}\{\lambda_1,\lambda_2,\ldots, \lambda_N\}$ includes the eigenvalues of the channel covariance matrix in descending order such that $\lambda_1 \geq \ldots \geq \lambda_N$. We also note that the size of matrices $\Sro^{'}\in \mathcal{C}^{K_a^\prime \times K_a^\prime}$ and $\Sro^{"} \in \mathcal{C}^{(N-K_a^\prime) \times (N-K_a^\prime)}$ depends on $K_a^{\prime}$, i.e., the number of bins which is filled by water. Based on the water-filling algorithm in Subsection \ref{Analog Beamformer Design Journal}, we readily conclude that
\begin{align}
\label{PD}
\Ueo&=\Uro
  \left[ {\begin{array}{cc}
   \sqrt{\frac{\nu}{\tp P_p}}\mathbf{I}_{K_{a^\prime}} & \mathbf{0} \\
    \mathbf{0} & \Sro^{''} \\
  \end{array} } \right]\Uro^H,	
	\end{align}
	\begin{align}
	\label{Semi PD}
\Uo&=\Uro
  \left[ {\begin{array}{cc}
   \Sro^{'}-\sqrt{\frac{\nu}{\tp P_p}}\mathbf{I}_{K_{a^\prime}} & \mathbf{0} \\
    \mathbf{0} &  \mathbf{0}\\
  \end{array} } \right]\Uro^H.
\end{align}         
Thus, after some manipulations we can complete the proof as
\begin{align} 
\mathrm{Tr}\big(\Uo \Ueo \big)&=\sqrt{\frac{\nu}{\tp P_p}}\mathrm{Tr}\bigg(\Uro\Big(\Sro^{'}-\sqrt{\frac{\nu}{\tp P_p}}\mathbf{I}_{K_{a^\prime}}\Big)\Uro^H \bigg) \nonumber \\
&=\sqrt{\frac{\nu}{\tp P_p}}\mathrm{Tr}\big(\Uo\big).
\end{align}
\vspace{-0.5cm}
\section{Proof of Proposition \ref{MMSE_error}}
\label{MMSE_error_ap}
By recalling (\ref{PD}) from Appendix \ref{Projection}, we can obtain the estimation error as 
\begin{align}
\mathrm{Tr}\big(\Ueo\big)&=\sqrt{\frac{\nu_{1}}{\tp P_p}}K'_a+\mathrm{Tr}\big(\Sro^{''}\big) \nonumber \\ 
&\leq \frac{K'_{a}}{\tp P_p }+\mathrm{Tr}\big(\Ro\big)-\mathrm{Tr}\big(\Sro^{'}\big)
\end{align}
where we used the fact that $\sqrt{\frac{\nu_{1}}{\tp P_p}}=\frac{K'_a}{\tp P_p K_a + \sum_{i=1}^{K'_a}\gamma_i}$ is upper bounded by $\frac{1}{\tp P_p}$. Now, in order to get deeper insights into how the performance of the estimator is affected by the channel covariance matrix, we normalize the estimation error in the following manner 
\begin{align}
\frac{\mathrm{Tr}\big(\Ueo\big)}{\mathrm{Tr}\big(\Ro\big)} \leq \frac{K'_a}{\tp P_p \mathrm{Tr}\big(\Ro\big)} + 1 -\frac{\mathrm{Tr}\big(\Sro^{'}\big)}{\mathrm{Tr}\big(\Ro\big)}.
\end{align}
This result reveals that the performance of the estimator can be improved if the ratio $\frac{\mathrm{Tr}\big(\Sro^{'}\big)}{\mathrm{Tr}\big(\Ro\big)}$ obtains higher values. In other words, a higher channel correlation will reduce the channel estimation error since the power of all eigenvalues is mostly concentrated into the strongest eigenvalues, i.e.  $\Sro^{'}$.
\bibliographystyle{IEEEtran}
\bibliography{IEEEabrv,refs}
\vskip -2\baselineskip plus -1fil

\begin{IEEEbiography}[{\includegraphics[width=1in,height=1.25in,clip,keepaspectratio]{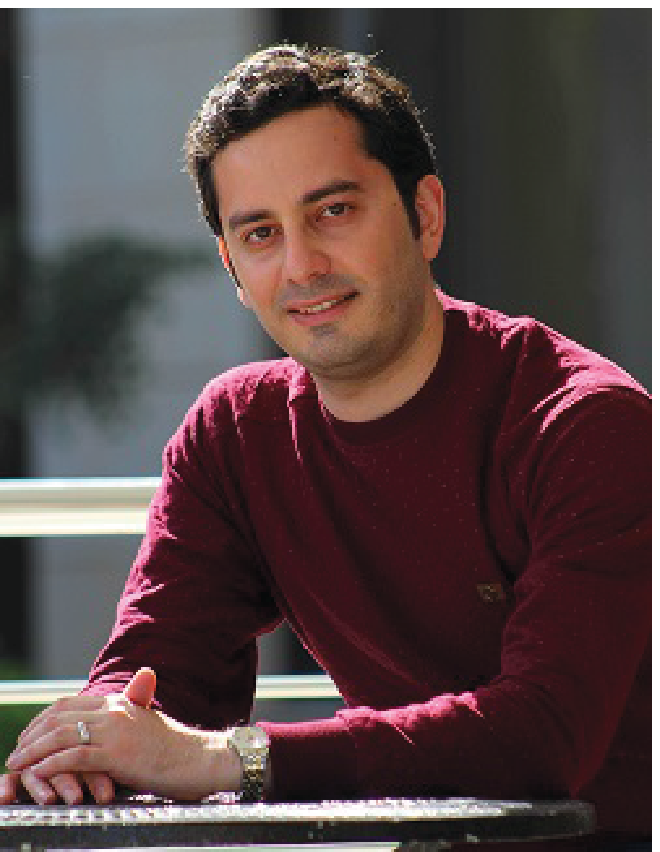}}]{Milad Fozooni} received his B.Sc. and M.Sc. degrees in electrical engineering from K. N. Toosi University of Technology and University of Tehran, Tehran, Iran in 2008 and 2012, respectively. From 2014 to 2017, he was a Ph.D. student at Queen's University Belfast where he particularly worked on ``Low-cost Architectures for Future MIMO Systems". Dr. Fozooni currently holds the position of Researcher in radio group at Ericsson Research, G\"oteborg, Sweden.  His main research interests include MIMO, massive MIMO systems, mmWave communications, signal processing and resource allocation techniques.
\end{IEEEbiography}

\vskip -1\baselineskip plus -1fil

\begin{IEEEbiography}[{\includegraphics[width=1in,height=1.25in,clip,keepaspectratio]{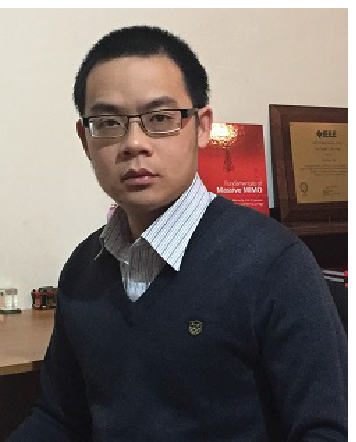}}]
{Hien Quoc Ngo}  received the B.S. degree in electrical engineering from the Ho Chi Minh City University of Technology, Vietnam, in 2007, the M.S. degree in electronics and radio engineering from Kyung Hee University, South Korea, in 2010, and the Ph.D. degree in communication systems from Link\"oping University (LiU), Sweden, in 2015. In 2014, he visited the Nokia Bell Labs, Murray Hill, New Jersey, USA. From January 2016 to April 2017, Hien Quoc Ngo was a VR researcher at the Department of Electrical Engineering (ISY), LiU. He was also a Visiting Research Fellow at the School of Electronics, Electrical Engineering and Computer Science, Queen's University Belfast, UK, funded by the Swedish Research Council.%
Hien Quoc Ngo is currently a Lecturer at Queen's University Belfast, UK. His main research interests include massive (large-scale) MIMO systems, cell-free massive MIMO, physical layer security, and cooperative communications. He has co-authored many research papers in wireless communications and co-authored the Cambridge University Press textbook \emph{Fundamentals of Massive MIMO} (2016).

Dr. Hien Quoc Ngo received the IEEE ComSoc Stephen O. Rice Prize in Communications Theory in 2015, the IEEE ComSoc Leonard G. Abraham Prize in 2017, and the Best PhD Award from EURASIP in 2018. He also received the IEEE Sweden VT-COM-IT Joint Chapter Best Student Journal Paper Award in 2015. He was an \emph{IEEE Communications Letters} exemplary reviewer for 2014, an \emph{IEEE Transactions on Communications} exemplary reviewer for 2015, and an \emph{IEEE Wireless Communications Letters} exemplary reviewer for 2016.
Dr. Hien Quoc Ngo currently serves as an Editor for the IEEE Wireless Communications Letters, Digital Signal Processing, and REV Journal on Electronics and Communications. He was a Guest Editor of IET Communications, special issue on ``Recent Advances on 5G Communications'' and a Guest Editor of  IEEE Access, special issue on ``Modelling, Analysis, and Design of 5G Ultra-Dense Networks'', in 2017. He has been a member of Technical Program Committees for several IEEE conferences such as ICC, GLOBECOM, WCNC, and VTC.
\end{IEEEbiography}

\vskip 0pt plus -1fil
\vspace{-1cm}
\begin{IEEEbiography}[{\includegraphics[width=1in,height=1.25in,clip,keepaspectratio]{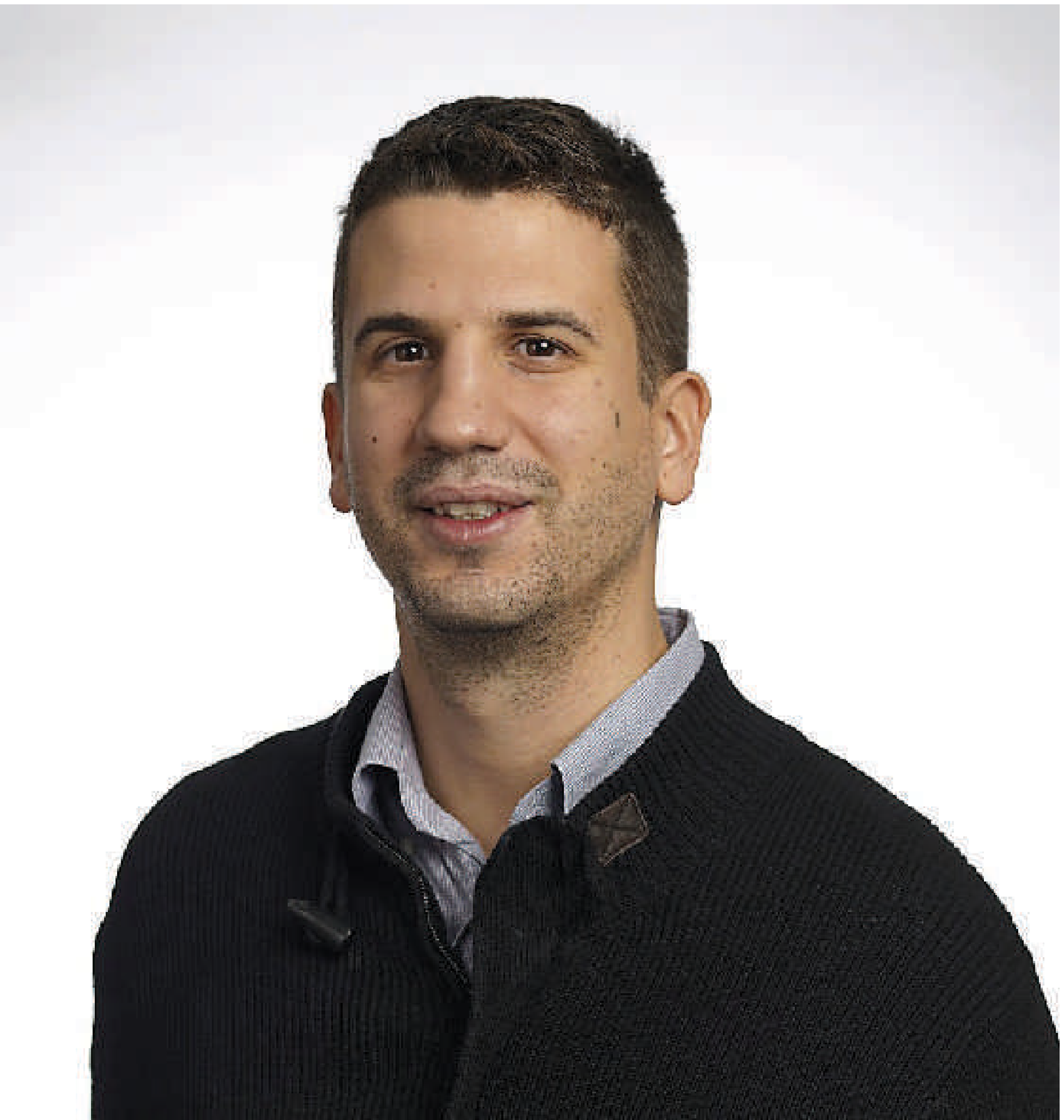}}]
{Michail Matthaiou}(S'05--M'08--SM'13) was born in Thessaloniki, Greece in 1981. He obtained the Diploma degree (5 years) in Electrical and Computer Engineering from the Aristotle University of Thessaloniki, Greece in 2004. He then received the M.Sc. (with distinction) in Communication Systems and Signal Processing from the University of Bristol, U.K. and Ph.D. degrees from the University of Edinburgh, U.K. in 2005 and 2008, respectively. From September 2008 through May 2010, he was with the Institute for Circuit Theory and Signal Processing, Munich University of Technology (TUM), Germany working as a Postdoctoral Research Associate. He is currently a Reader (equivalent to Associate Professor) in Multiple-Antenna Systems at Queen's University Belfast, U.K. after holding an Assistant Professor position at Chalmers University of Technology, Sweden. His research interests span signal processing for wireless communications, massive MIMO, hardware-constrained communications, and performance analysis of fading channels.

Dr. Matthaiou and his coauthors received the 2017 IEEE Communications Society Leonard G. Abraham Prize. He was the recipient of the 2011 IEEE ComSoc Best Young Researcher Award for the Europe, Middle East and Africa Region and a co-recipient of the 2006 IEEE Communications Chapter Project Prize for the best M.Sc. dissertation in the area of communications. He was co-recipient of the Best Paper Award at the 2014 IEEE International Conference on Communications (ICC) and was an Exemplary Reviewer for \textsc{IEEE Communications Letters} for 2010.
\end{IEEEbiography}
\break
\begin{IEEEbiographynophoto}{\relax}\unskip
In 2014, he received the Research Fund for International Young Scientists from the National Natural Science Foundation of China. In the past, he was an Associate Editor for the \textsc{IEEE Transactions on Communications}, Associate Editor/Senior Editor for \textsc{IEEE Communications Letters} and was the Lead Guest Editor of the special issue on ``Large-scale multiple antenna wireless systems" of the \textsc{IEEE Journal on Selected Areas in Communications}. He was a co-chair of the Wireless Communications Symposium (WCS) at IEEE GLOBECOM 2016. 
\end{IEEEbiographynophoto}

\vskip 0pt plus -1fil
\vspace{-0.5cm}
\begin{IEEEbiography}[{\includegraphics[width=1in,height=1.25in,clip,keepaspectratio]{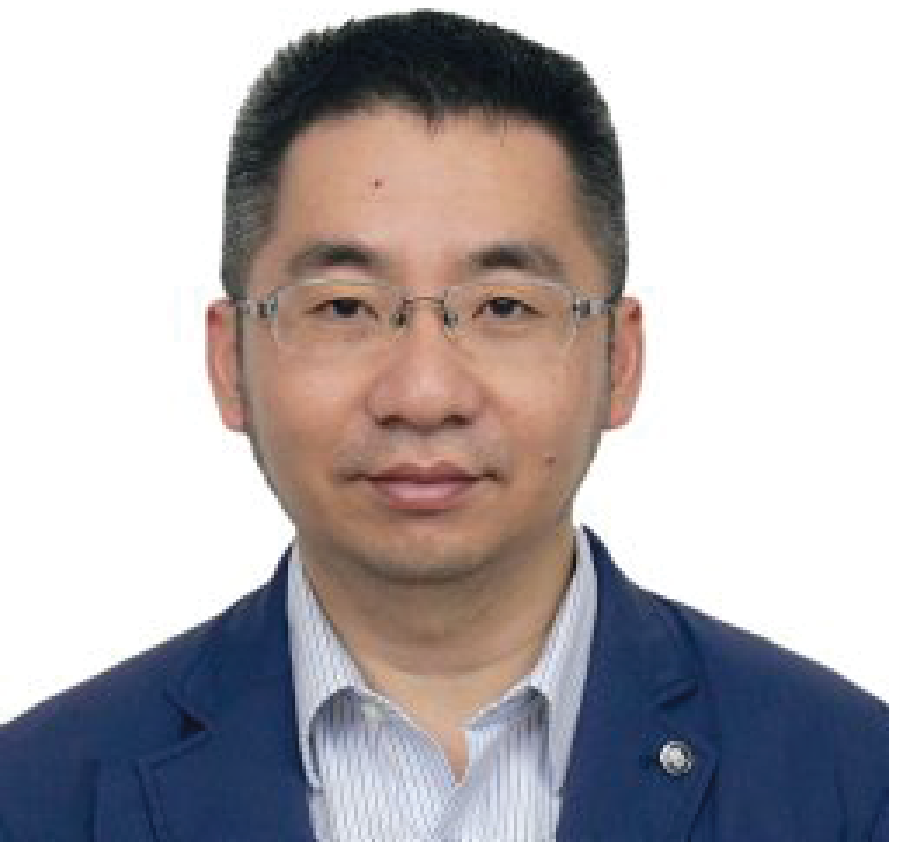}}]{Shi Jin} (S'06--M'07--SM'17) received the B.S. degree in communications engineering from Guilin University of Electronic Technology, Guilin, China, in 1996, the M.S. degree from Nanjing University of Posts and Telecommunications, Nanjing, China, in 2003, and the Ph.D. degree in information and communications engineering from the Southeast University, Nanjing, in 2007. From June 2007 to October 2009, he was a Research Fellow with the Adastral Park Research Campus, University College London, London, U.K. He is currently with the faculty of the National Mobile Communications Research Laboratory, Southeast University. His research interests include space time wireless communications, random matrix theory, and information theory. He serves as an Associate Editor for the  \textsc{IEEE Transactions on Wireless Communications}, and  \textsc{IEEE Communications Letters}, and  \textsc{IET Communications}. Dr. Jin and his coauthors have been awarded the 2011 IEEE Communications Society Stephen O. Rice Prize Paper Award in the field of communication theory and a 2010 Young Author Best Paper Award by the IEEE Signal Processing Society.
\end{IEEEbiography}

\vskip 0pt plus -1fil
\vspace{-0.5cm}
\begin{IEEEbiography}[{\includegraphics[width=1in,height=1.25in,clip,keepaspectratio]{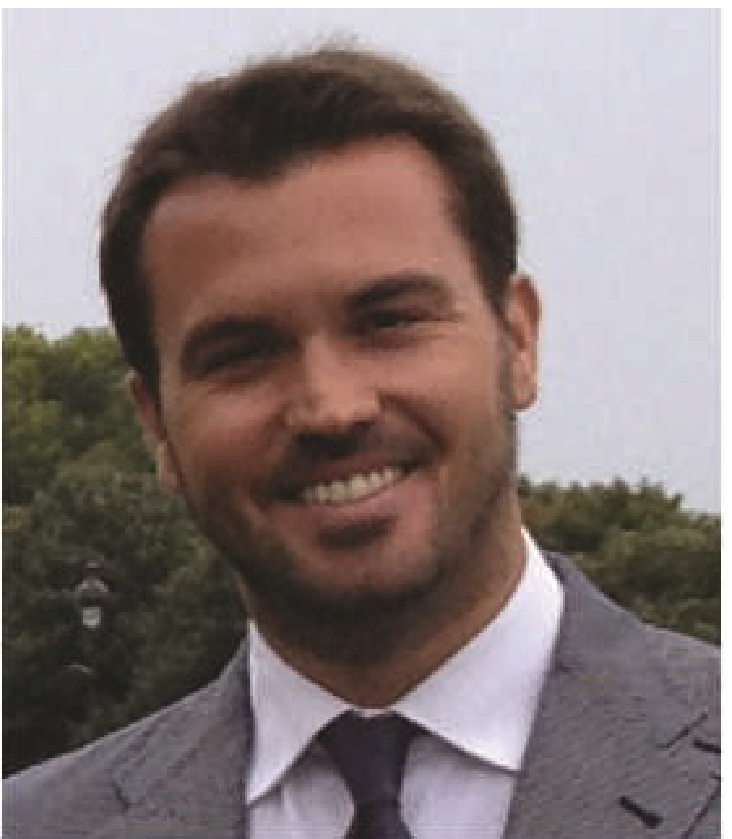}}]{George C. Alexandropoulos} (S'07--M'10--SM'15) was born in Athens, Greece, in 1980. He received the Engineering Diploma degree in computer engineering and informatics, the M.A.Sc. degree (with distinction) in signal processing and communications, and the Ph.D. degree in wireless communications from the University of Patras (UoP), Rio-Patras, Greece, in 2003, 2005, and 2010, respectively. From 2001 to 2010, he has been a Research Fellow with the Signal Processing and Communications Laboratory, Department of Computer Engineering and Informatics, School of Engineering, UoP. From 2006 to 2010, he was with the Wireless Communications Laboratory, Institute of Informatics and Telecommunications, National Center for Scientific Research ``Demokritos," Athens, Greece, as a Ph.D. Scholar. From 2007 to 2011, he has been collaborating with the Institute for Astronomy, Astrophysics, Space Applications, and Remote Sensing, National Observatory of Athens, Greece, where he participated
in one national and two European projects. Within 2011, he also worked with the Telecommunication Systems Research Institute, Technical
University of Crete, Chania, Greece, in the framework of one European project. In the summer semester of 2011, he was an Adjunct Lecturer with the Department of Telecommunications Science and Technology, University of Peloponnese, Tripoli, Greece. From 2011 to 2014, he was a Senior Researcher with the Athens Information Technology Center for Research and Education, where he has been involved with the technical management of four European projects and lectured several mathematics courses. Since 2014, he has been a Senior Research Engineer with the Mathematical and Algorithmic Sciences Laboratory, Paris Research Center, Huawei Technologies France SASU, Boulogne-Billancourt. His research interests lie in the general areas of performance analysis and signal processing for wireless networks, with emphasis on multi-antenna systems, interference management, high-frequency communication, cooperative networking, energy harvesting, and cognitive
radios. 

Dr. Alexandropoulos is a Senior Member of the IEEE Communications and Signal Processing Societies as well as a Professional Engineer of
the Technical Chamber of Greece. He currently serves as an Editor for  \textsc{IEEE Transactions on Wireless Communications} and \textsc{IEEE Communications Letters}. He received a Postgraduate Scholarship from the Operational Programme for Education and Initial Vocational Training II, Ministry of Education, Lifelong Learning, and Religious Affairs, Republic of Greece; a student travel grant for the 2010 IEEE Global Telecommunications Conference in Miami, USA; and the Best Ph.D. Thesis Award by a Greek University in the fields of informatics and telecommunications from the Informatics and Telematics Institute, Thessaloniki, Greece, in 2010.
\end{IEEEbiography}

\end{document}